# Developing A Personal Decision Support Tool for Hospital Capacity Assessment and Querying

Robert L Burdett[1,3], r.burdett@qut.edu.au
Paul Corry[1], p.corry@qut.edu.au
David Cook[2], d.cook@uq.edu.au
Prasad Yarlagadda[3], y.prasad@qut.edu.au

[1] School of Mathematical Sciences, Queensland University of Technology, GPO Box 2434, 2 George Street Brisbane Qld 4000, Australia
[2] Princess Alexandra Hospital, 2 Ipswich Rd, Woolloongabba, Brisbane, Qld 4102, Australia
[3]School of Mechanical, Medical & Process Engineering, Queensland University of Technology, GPO Box 2434, 2 George Street Brisbane Qld 4000, Australia

**Abstract:** This article showcases a personal decision support tool (PDST) called HOPLITE, for performing insightful and actionable quantitative assessments of hospital capacity, to support hospital planners and health care managers. The tool is user-friendly and intuitive, automates tasks, provides instant reporting, and is extensible. It has been developed as an Excel Visual Basic for Applications (VBA) due to its perceived ease of deployment, ease of use, Office's vast installed userbase, and extensive legacy in business. The methodology developed in this article bridges the gap between mathematical theory and practice, which our inference suggests, has restricted the uptake and or development of advanced hospital planning tools and software. To the best of our knowledge, no personal decision support tool (PDST) has yet been created and installed within any existing hospital IT systems, to perform the aforementioned tasks. This article demonstrates that the development of a PDST for hospitals is viable and that optimization methods can be embedded quite simply at no cost. The results of extensive development and testing indicate that HOPLITE can automate many nuanced tasks. Furthermore, there are few limitations and only minor scalability issues with the application of free to use optimization software. The functionality that HOPLITE provides may make it easier to calibrate hospitals strategically and/or tactically to demands. It may give hospitals more control over their case-mix and their resources, helping them to operate more proactively and more efficiently.

## 1. Introduction

Hospitals play a vital role in health care systems worldwide. Operating continuously all year-round, most hospitals have the capability to treat and care for a diverse cohort (i.e., group) of patients with different illnesses and conditions, and with different health care requirements. Yet the goal of delivering quality care to as many patients as possible, at an affordable cost, is an ongoing challenge (Krueger (2018)). Hospital output is restricted by finite and often dissimilar health care resources, and these are limited by diminishing health care funding (Walczak et.al. (2002)). The inherent conflicts between quality, cost, and output, are ever present, and require careful and constant mediation (Burdett and Kozan (2016)).

There is an essential need for a holistic and predictive view of an entire hospital to understand the factors driving outputs and financial performance (Kreuger (2018)). Currently most hospitals function in a reactive manner and have a short-term insular focus. Based upon staffing levels and the current setup of theatres, wards, beds, and so forth, patients are chosen from the waiting list and treated opportunistically. However, this often leads to significant imbalances between the supply and demand for medical resources (Emanual et. al. 2020). To provide a long-term remedy for this issue it is necessary to fully understand how various structural and operational factors affect a hospital's performance and output. For instance, there is a need to quickly provide situational awareness around the prioritisation, allocation and sharing of hospital resources and the performance impacts of capacity-related decisions like adding or removing beds in wards, building new theatres, changing the master surgical schedule (i.e., adding or removing sessions, changing session duration or sessions per day, changing the number of days theatres operate per week), treating new patient types, applying new medical or surgical techniques, and so forth.

To provide a comprehensive view of hospital capacity this article proposes the development of a *software solution* with embedded optimisation methods to help hospital planners and staff perform insightful quantitative assessments of hospital capacity and utilization. Determining whether an adequate approach can be devised and put into practice is our primary line of inquiry. How best to facilitate the integration of mathematical techniques is our second line of enquiry. This articles' approach builds upon the research presented Burdett and Kozan (2016) and Burdett et. al. (2017) that is now well established in the literature. In those articles, mathematical models and other quantitative techniques were successfully developed to perform holistic hospital capacity assessment and capacity querying activities. A case study of a large tertiary hospital was previously used to validate those approaches. The practicalities of designing an appropriate decision support tool and integrating those models, however, was not considered, and provides the motivation for this article.

Our software solution is called HOPLITE, which stands for hospital planning, intel, and tactical evaluation. To the best of our knowledge our software solution, is a new capability, that can replace the inexact ad-hoc calculations often performed by hand by health care managers and planners. HOPLITE is a prototype and a minimum viable product for performing various capacity assessment and capacity querying tasks. HOPLITE also provides the capability to check whether a hospitals' configuration, layout and resources are sufficient to meet current and future demands. It also suggests outputs which are achievable when demands cannot be met.

A foremost aim of the HOPLITE software is to handle the many nuances that make "prompt" capacity assessments troublesome. The software and its' graphical user interfaces have been implemented in Microsoft Excel using VBA and can be run on any personal computer. Excel has been chosen as the driver of the PDST because it is an event driven framework, and because some capacity assessment tasks and queries require a linear and non-linear programming solver. The PDST makes uses of Excels' inbuilt optimization solver or the more capable and unrestricted "OpenSolver" add-in (Mason (2012)) so that hospitals do not have to purchase a licence or subscription for more capable optimization software like IBM CPLEX or GUROBI, nor consider the integration of a PDST with those software packages. The tool is regarded as a PDST because it is developed for one manager and/or planner or a small team.

The format of this article is as follows. In Section 2 a literature review and analysis are provided. In Section 3 the details of the quantitative framework and techniques used in HOPLITE are provided commencing with an outline of key technical details. The specification, capabilities and graphical user interfaces are then presented in Section 4. Design strategies employed during development are also provided and examples of how the PDST is used to perform various assessments is shown. Conclusions and final remarks are given in Section 5. Broader issues including potential further extensions to the software are also discussed.

## 2. Literature Review

In this section, techniques for hospital case mix planning, capacity assessment and capacity allocation are reviewed, and the current state of the art is described. Some other loosely related topics like operating room planning/scheduling with a focus on capacity are also included. Our discussion describes the decision problem addressed, and the methods used to solve it. Interface development and the integration of mathematical techniques is then focussed upon.

## 2.1. Hospital Capacity and Case Mix Planning

Hospital case-mix planning, and capacity assessment are contemporary topics. In recent times there has been considerable interest from researchers, academics, and other decision makers. Deficiencies in existing health care systems and practices, exacerbated by the COVID pandemic, has fuelled research to find better ways to plan and manage health care resources. In past research, a variety of approaches have been applied to the aforesaid decision problem, including mixed integer programming (Ma et al. (2011), Burdett et al. (2017), Shafaei and Mozdgir (2018)), stochastic programming (Neyshabouri and Berg (2017), Freeman et al. (2018), McRae and Brunner (2019), Burdett et al. (2022)), process mining (Andrews at al. (2022)) and multicriteria optimisation (Malik et al. (2015), Burdett and Kozan (2016), Zhou et al. (2018)). Table 1 summarises the most crucial details about recent research.

**Table 1.** Summary of recent CMP and related research

Abbreviation: Case Mix Scheduling (CMS); Case Mix Planning (CMP); Discrete Event Simulation (DES); Epsilon Constraint Method (ECM); Graphical User Interface (GUI), Hospital Capacity Assessment (HCA); Linear Programming (LP); Multicriteria (MC); Non-Linear Programming (NLP); Operating Room Planning (ORP); Operating Room Scheduling (ORS); Regional (REG); Sample Avg. Appr (SAA);

| Article | Problem | ED | OR | WARD | ICU | Objectives | STOCH | MC | GUI | REG | Method |
|---|---|---|---|---|---|---|---|---|---|---|---|
| Ma et al. (2011) | CMS | × | ✓ | ✓ | × | Profit | × | × | × | × | MIP |
| Ma & Demeulemeester (2013) | CMP-ORS | × | ✓ | ✓ | × | Profit, Bed Shortage | × | × | × | × | MIP |
| Malik et.al. (2015) | ORP | × | ✓ | × | × | Waiting List Size, Costs | × | ✓ | × | × | Meta H. |
| Jebali and Diabat (2015) | ORP | × | ✓ | ✓ | ✓ | Costs | ✓ | × | × | × | SAA |
| Burdett & Kozan (2016) | HCA | × | ✓ | ✓ | ✓ | Output × 21 | × | ✓ | × | × | LP, ECM. |
| Yahia et al. (2016) | CMP | × | ✓ | ✓ | ✓ | Output | ✓ | × | × | × | SAA |
| Jebali and Diabat (2017) | ORP | × | ✓ | × | ✓ | Cost | ✓ | × | × | × | SAA |
| Burdett et al. (2017) | HCA | ✓ | ✓ | ✓ | ✓ | Output | × | × | × | × | LP |
| Zhou et al. (2018) | HCA | × | ✓ | ✓ | × | Revenue, Equity | ✓ | ✓ | × | × | DES, MIP, ECM |
| Shafaei & Mozdgir (2018) | ORP | × | ✓ | ✓ | ✓ | Value | ✓ | × | × | × | LP & TOPSIS |
| Freeman et al. (2018) | ORS | × | ✓ | ✓ | ✓ | Payment | ✓ | × | × | × | MIP |
| McRae et al. (2018) | CMP | × | ✓ | ✓ | ✓ | Profit | × | × | × | × | NLP |
| McRae & Brunner (2019) | CMP | × | ✓ | ✓ | ✓ | Revenue | ✓ | × | × | × | SAA |
| Burdett et al. (2022) | HCA | × | ✓ | ✓ | ✓ | Output, Unmet Demand, Outsourcing | × | ✓ | × | ✓ | MIP |
| Saha & Rathore (2022) | CMP | × | × | × | ✓ | Expected Cost | ✓ | × | × | × | Heuristic |
| This article | HCA | × | ✓ | ✓ | ✓ | Output | × | × | ✓ | × | MIP |

In summary, Ma et al. (2011) developed and tested a case mix planning model maximizing the overall financial contribution of a hospital. Chen et al (2015) developed quantitative approaches for patient flow scheduling and capacity planning in a rheumatology department, which would be useful in a smart hospital and health care environment. Malik et.al. (2015) formulated and solved a bi-objective aggregate capacity planning problem for operating theatres. Yahia et al. (2016) applied the sample average approximation (SAA) approach to solve a stochastic planning model. They considered the selection of a case mix for a single surgical department, with uncertain surgery durations, length of stay and demand. The number of theatre hours assigned to each patient group was a primary decision. Burdett and Kozan (2016) and Burdett et. al. (2017) developed the first holistic hospital capacity allocation approach. They include case mix constraints in their deterministic mixed integer programming (MIP) model or else impose multiple objectives, from which a set of non-dominated capacity solutions can be generated. Their approach has been applied to a 21 objective real-life scenario. Burdett et al. (2022) provided the first regional hospital capacity allocation approach and applied it to a 15-hospital regional case study. Zhou et.al. (2017) considered the capacity allocation of hospital wards and the joint optimization of hospital revenue and equity among several types of patients. In response they proposed a multi-objective stochastic programming model with two objectives. As their objective functions have no "closed form" they used a data-driven discrete-event simulation to evaluate random patient arrivals and lengths of stay. An adaptive epsilon-constraint algorithm (ECM) and a multi-objective Genetic algorithm were developed to solve the proposed non-linear mathematical model. Shafaei and Mozdgir (2018) developed a mathematical model to optimize the allocation of OR time among surgical groups and applied a robust estimator for values of the model parameters. McRae, Brunner, Bard (2018) developed a non-linear mixed-integer programming model and incorporated economies of scale. Freeman et al. (2018) considered case mix planning and developed a multi-phase approach to generate a set of candidate solutions. They applied simulation techniques to evaluate the master surgical schedule (MSS) and each case mix solution. McRae and Brunner (2019) presented a framework for evaluating the effect of stochastic parameters on the case mix of a hospital. Liu et al. (2019) developed for the daily scheduling of surgical patients, an integrated scheduling and capacity planning approach. They declare that "traditional scheduling policy, driven by operating room usage, may lead to significantly suboptimal use of downstream capacity and may result in up to a three-fold increase in total expenses". In contrast, "a scheduling policy based on downstream capacity usage often performs close to an integrated scheduling policy, and therefore may serve as a simple, effective scheduling heuristic for hospital managers—especially when the downstream capacity is costly and less flexible". Burdett et al. (2022) provided the first regional HCA approach and applied it to a 15-hospital regional case study. Saha and Rathore (2022) considered physicians as a significant limiting factor in hospital care. They developed a two-stage stochastic programming approach in which decisions on regular physician allocation and their capacity adjustments are a trade-off between expected cost and patient demand fulfillment. To solve the problem inexactly, a scenario-based heuristic was applied with 1000 scenarios.

***Information Technology and Decision Support Systems.*** Around the world there are various IT platforms for decision making. In the literature the following are prevalent: decision support systems (DSS), expert systems (ES), executive information system (EIS), management information systems (MIS), and management support systems (MSS). There does not appear to be a distinct boundary or delineation between the different types. As far as we can pertain, a management information system organizes and retrieves data and generates reports that summarize activities and performance of interest to managers (Burstein and Holsapple (2008)); however, an expert system is akin to an electronic counsellor, delivering expertise to appropriate staff (Forgionne and Kohli (1996)).

Of the different types of IT platform developed, personal decision support systems (PDSS) stand out. PDSS were very well-used in the 1980s and are a pure form of IT-based management support, generally thought of as superior to MIS (Arnott (2008)). The emphasis of PDSS is to empower and support individual managers. The term PDSS is not well known or used nowadays and has been replaced with the general term analytics (Arnott (2008)). What makes PDSS development projects different to others is that users do not know what they want, and analysts do not understand what users need, the analyst and user cannot provide functional specifications, the users' concept of the task will be shaped by the DSS, and users have the autonomy to tackle the tasks in a variety of ways. The term PDSS best fits the type of software developed in this article, and the project environment upon which it has been developed.

Regarding health care applications, the following systems are popular: clinical decision support system (CDSS), hospital information system (HIS), healthcare information technologies (HIT), and hospital management support systems (HMSS). Computerized CDSS have rapidly evolved since the 1980s (Sutton et.al. 2020). This specific type of management information system is specifically designed to aid clinical decision making by promptly providing actions, advice, alerts, and reminders, to a clinician. CDSS software often matches individual patients to a computerized clinical knowledge base. They are either knowledge-based or non-knowledge based. In the former type, rules are programmed, if-then actions are created, and expert medical knowledge is followed. In the later, artificial intelligence, machine learning and other statistical methods are applied.  In recent years Rahimi et.al. (2016) developed a dynamic risk-based framework for patient prioritization. This is regarded as a complex decision-making process, currently skewed for instance by surgeons' opinions, and a static assessment of a patients' condition. That article highlights the development of user-friendly software as a necessary development task to facilitate the implementation of the proposed framework. Sutton et.al. (2020) analysed the application of CDSS and identified both positives and negatives. The biggest downsides include, maintaining necessary databases, the complexity of data integration, keeping up with changes in data, the use of poor-quality data, the use of what is regarded as sensitive information, and financial viability to develop and setup new CDSS systems. Advantages include instant access to patient-specific information, the provision of assessments and recommendations, the capability to send reminders for preventative care, and alerts about potentially dangerous situations.

Forgionne and Kohli (1996) are one of the first to propose the development and application of hospital management support systems (HMSS). They view hospitals as a "make-to-order" enterprises and use concurrent engineering (CE) principles in their HMSS. The goal of their HMSS is to improve quality, reduce costs, and decrease the lead time from admission to discharge for new or readmitted patients. James et.al. (2010) describes the development of a DST for expert elicitation of data. The integration of sophisticated mathematical techniques is focused upon. They demonstrate how to create a viable DST in Java, using object-oriented design, open-source libraries, data persistence using MySQL, and the application of the R statistical software to perform all statistical calculations. The software architecture is based around the creation of a project model that serves to encapsulate all data and data relationships necessary for an elicitation project. Each project consists of several independent project phases. Gupta and Sharda (2013) and Bardhan and Thouin (2013) have reviewed the application and current state of research of healthcare information technologies and the application of informatics. Six key themes have been identified in the former article. In the later article, it was identified that spending on health IT does matter, and this has important policy implications for investments in health IT. Additionally, the usage of financial management systems is associated with lower hospital operating expenses. Sebaa et. al. (2017) developed a medical decision support platform and clinical relational database as a medical data warehouse. The platform is intended to identify health care trends and to report other important statistics to users. Bodina et.al. (2017) considered the identification of objective criterions for the strategic management of a complex hospital and the allocation of scarce hospital resources. An evaluation (a.k.a., scoring) system was described based upon six themes, namely strategic, operating, research, economic, organizational, and quality. No actual "concrete" mathematical methods were however provided or tested. Bruggemann et.al. (2021) developed a web accessible simulation-based decision support tool to explore hospital resource usage in high demand circumstances like the COVID pandemic. They advocate the development of "what-if" scenarios for the evaluation of stochasticity and the export of results to a CSV file to enable connectivity to other software suited. The capability to provide meaningful and easy to understand outputs is also emphasised.

***Corporate Solutions.*** Corporate healthcare analytics software and services is increasing and there are numerous market drivers. The need for better capacity management and pricing in hospitals and a need to curtail healthcare costs are explicitly quoted in an online report (Hospital Capacity Management Solutions Market). In that report, healthcare software is described as either integrated or standalone, and delivered on-premises, or by the cloud. The following products are described as most popular at present:

- Asset Management (Medical Equipment Management, Bed Management)
- Patient Flow Management Solutions
- Workforce Management (Nursing & Staff Scheduling Solutions, Leave and Absence Management)
- Quality Patient Care

The major players in the global hospital capacity management solutions market are listed as Cerner Corporation (US), McKesson Corporation (US), HealthStream (US), Stanley Healthcare (US), and Halma plc (US). Other prominent players in this market include Infosys (India), Teletracking Technologies, Inc. (US), NextGen Healthcare (US), Allscripts Healthcare Solutions, Inc. (US), Epic Systems Corporation (US), Sonitor Technologies (US), Koninklijke Philips N.V. (Netherlands), Neusoft Corporation (China), Infinitt Healthcare Co., Ltd. (South Korea), JVS Group (India), Infor Systems (US), Care Logistics (US), WellSky (US), Simul8 Corporation (US), and Alcidion Corporation (Australia).

Dashboards have become particularly popular, and many are produced and packaged in business intelligence solutions (https://www.sisense.com). Most dashboards summarise outcomes, performance, and cost information over a given period.

***General Findings.*** Hospital case mix planning and capacity allocation/ assessment is a niche research area with relatively few articles overall. Despite the arguments concerning the efficacy and potential of recent analytical approaches, actual decision support tools (DST) using those approaches have not been described. From this we infer that they have not been implemented and deployed, and if they have, it is not extensive. We also infer that prior articles on the topic are far too nuanced for hospital planners, information technology staff, and managers, to understand and interpret. As such there is little motivation to progress these methods to the point where they can be deployed within hospital information systems.

Providing advanced predictive analytics is an important goal in the healthcare industry (Kreuger (2018)). The development of planning software is, however, challenging. To apply a software solution, it is necessary to fully understand the needs of the potential hospital users and to ensure access to appropriate hospital information. This is also true of analytical methods. Unfortunately, hospitals are data rich, and information poor (Adeyemi et. al. (2013)). Raw data is frequently held hostage in disparate enterprise IT systems, data archives and data warehouses (Kreuger (2018)) and access may be granted to only a select number of personnel within an organisation. The time to extract information from raw data stored within the IT system may also be prohibitive and is at the very least, nuanced.

Despite the great need for decision support tools, there appears to be a significant lack of them to implement and put into practice (Humphreys et al. 2022). Of the existing decision support tools found in the literature, anecdotal evidence suggests very few hospitals have adopted them for regular use. As Hulshof et al. (2013) reports, current techniques are too limited, being too myopic, focussing only on the development of long-term cyclical plans, and are incapable of providing solutions for real-life sized instances. Hawkinson et al. (2018) also comments that integrated models that can tie together competing metrics in capacity planning decisions are not being developed. Nor are there tools that provide sufficient “what-if” capabilities to support managerial decision making.

For decision making and planning it is vital that hospital personnel transform information into task, event, and process knowledge. According to Forgionne and Kohli (1996), comprehensive decision support requires consolidation of separate information systems or data sources, and the effective delivery of integrated capabilities and knowledge, involving clinical and administrative information, in a systematic, complete, and timely manner. However, all these activities are still complex to this day, and it would appear, there are no easy shortcuts in the development of effective DSS and DST.

Information technologies are often described as having the potential to improve both the quality and effectiveness of healthcare providers (Bardhan and Thouin (2013)). However, the impact of health information technologies (HIT) on healthcare delivery and quality of care is difficult if not impossible to determine, without first applying them for an extended period and comparing metrics of performance. The quality of the human user interface also plays an important part in the adoption of many information technology and decision support systems. The aesthetics, however, cannot justify lack of capability and essential analyses and evaluations must be present.

In the cited literature, various DSS and DST are described. However, most articles only provide very high-level explanations, and many important technical details are not described. There is truly little advice or transferrable methodology. Furthermore, there is no detailed description of why interfaces are set up a particular way and how alternative layouts affect decision making. At this point in time, very few of the DSTs proposed in health-related articles use and or integrate sophisticated mathematical techniques. Of the papers that do provide sophisticated mathematical techniques, very few consider how analytical techniques could be used by hospitals, and as such, only contribute to the literature. Overall, there are few managerial insights provided in those technical articles.

Many DSTs are described as tools to make health care resource allocations. Most of those, however, suggest a manual approach, not an automated optimization approach. We have observed a variety of claims that proposed data warehouses alone will help decision makers (i.e., like Sebaa et.al.), however, there is no evidence to suggest that decision makers would know how to use the data that is collected and archived in their platforms. Nor is there any evidence to suggest that by simply looking at dashboard results, it would be clear how to apportion resources optimally.

There are many DSS in other domains, and this leads us to the conclusion that those may be a good reference point for healthcare and should be investigated in more detail. For instance, the article by James et.al. (2010) is a model of how mathematical techniques can be implemented successfully in a DSS. A review of other domains, however, is a considerable task, outside the scope of this article.

## 3. Quantitative Framework

In this section, the mathematical framework behind HOPLITE is detailed. HOPLITE is a predictive analytics tool designed for health care managers (HCM). HCM have a variety of strategic and operational responsibilities, some of which are delegated to other staff. One of their key strategic responsibilities, however, is capacity planning (Ozcan (2017)). HOPLITE was developed and validated collaboratively with hospital administrators and clinicians who were involved in our research project.

### 3.1. General Overview and Assumptions

The terms and concepts below are important for understanding HOPLITE, and it is necessary to discuss them first.

***Hospital Capacity.*** Hospital capacity is viewed as the number of patients of different types treatable over time or the number of activities of different types that can be performed over time. As such, it can also be viewed as a rate.

***Patient Types.*** Hospitals treat and care for a diverse cohort of patients with very different illnesses and conditions, with very different health care requirements. For planning purposes, it is necessary to aggregate patients into a finite and more manageable number of patient type groupings. Aggregation by specialty (i.e., medical, and surgical) and by condition (i.e., diagnostic related group (DRG)) is logically appropriate based upon our observations and consultations with hospital practitioners. Other approaches could be taken, based for instance upon the clustering of patients with similar attributes, however, this does not conform with the way hospitals themselves classify patients, and would inevitably be less intuitive to the end users. Patients of the same type often have different treatments and resource requirements, and this necessitates the definition of a sub-group or sub-type.

**Hospital Activities**. The purpose of HOPLITE is to perform various assessments, evaluations and queries relating to hospital activity. In our prototype tool, typical high-level activities like pre-operative care (preop), surgery (sur), post anaesthesia care (pac), post-operative care (postop), and intensive care (ic) are currently included. Lower-level activities are not considered for the simple reason that they are of dubious value strategically and overcomplicate the assessments and the software. This, however, is not to say that they could and should not be included at some point, in another version. Preoperative care and post anaesthesia care occur either in a specific surgical care area or in a ward bed. Also, any significant surgery occurs in an operating theatre and post-operative recovery is almost always performed in a ward bed.

***Patient Pathways***. The path taken by patients during their hospital stay is an important piece of information relevant to an assessment of hospital capacity. Patient pathways can be used to distinguish patients of different types and sub types. Patient pathways can be defined in diverse ways. In this article a patient pathway is deemed a list of the places (a.k.a., hospital areas) a patient visits and a description of the activity(s) performed in each of those places. Formally, $pathway = \{(act, w, t) | w \in W, t \in \mathbb{R}\}$ where $W$ is the set of hospital areas and $t$ is the length of stay (a.k.a., occupation time) in that area, recorded in a suitable unit of measure like hours or minutes.

Pathways should be extracted for each patient type grouping $g \in G$. They can either be extracted from hospital records or inferred. Regarding surgical patients, there are a finite number of surgical pathways. We have observed that most surgical patients have a similar set of activities and a common flow through the hospital. The paths that may be taken is summarised in the directed "cyclic" graph shown in Figure 1. In this network there are no unknown entry and departure points. Assuming the possibility of at most two surgeries (i.e., two cycles), this network produces 51 paths. These can be extracted using a depth first traversal algorithm.

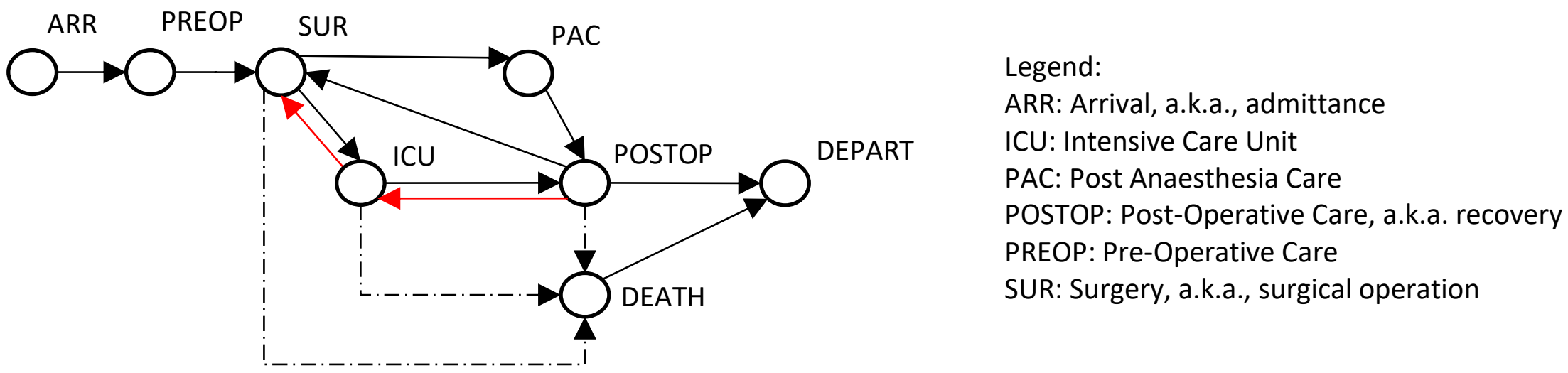

**Figure 1.** Network of potential pathways for surgical patients

The network in Figure 1 includes a death event activity. This is worth adding because patients that die during surgery, intensive care, and postop activities have a different "resource consumption" profile. If these events are not distinguished, expected activity durations will be in error. Any random variable created from the data will be "bi-modal" and may have unnecessarily long "tails".

***Patient Profiles.*** A resource consumption profile (RCP) is recommended to be created for each patient type or sub type. Conceptually it is a set of the resources required and the time each of those is required. A profile is formally defined as a set of tuples $\{(r, t) | r \in R, t \in \mathbb{R}\}$ where the time utilisation $t$ is measured in either minutes or hours. From a practical perspective, the time utilisation is unlikely to be invariable. As such it is worth computing an expected (a.k.a., averaged) value based upon historical or other empirical evidence. Not all resources need to be included; a guideline would be to include "principal" resources, for instance, which have been established as bottlenecks, as being scarce, or as being highly costly. A minimal profile may also include only locations. In many patient pathways the same location is visited numerous times, or the same resource is used for different activities. When assessing capacity, how many times a location or any other resource is required is not important, but the total time required is. In this article, only hospital areas are considered as resources. As such, the RCP is $\{(w, t) | w \in W, t \in \mathbb{R}\}$.

The RCP discussed above is "inherently linked" to the patient pathway discussed earlier. A patient pathway can be converted into a RCP by aggregating pathway records with the same location or resource. This can be done iteratively by adding tuples from the pathway to the profile.

**Example**. If $pathway = [(preop, sca, t_1),(sur, ot, t_2),(ic, icu, t_3),(postop, ward, t_4),$ $(preop, ward, t_5),(sur, ot, t_6),(pac, sca, t_7),(postop, ward, t_8)]$ then $profile = [(sca, t_1 + t_7), (ot, t_2 + t_6), (ward, t_4 + t_5 + t_8), (icu, t_3)]$.

**Patient Case Mix and Sub Case Mix.** A sub-division of a patient cohort into specific patient types is called the case mix. The case mix is typically viewed as either the proportion of each patient type or as an actual number of patients of each type. A sub-division of the patients of a particular type into different sub-types is called the sub case mix or simply sub mix. The sub mix is the proportion of each sub-type.

**Master Surgical Schedule (MSS):** How often theatres are used, and how effectively they are used, affects the output of a hospital. The operating theatres of a hospital operate on certain days of the week and at certain times during the day. This results in a set of theatre sessions, which can be used for surgical procedures. A master surgical schedule (MSS) is a plan that describes which hospital unit or specialty has access to each theatre session.

**Final Remarks:** The creation of this PDST has raised some key questions. How much detail should be included is one question of interest. It is hypothesised that a minimum viable PDST should only include theatre, ward, and intensive care unit areas. Similarly, only surgery, intensive care and postop activities should be included. Surgical care areas and the associated pre-operative and post anaesthesia care activities could be left out because the time requirements are small at those locations, and those areas usually have excess bed capacity. Individual treatment spaces, like beds, should not be treated independently; they should be aggregated. As such, task allocations to individual beds are avoided. This contrasts with the numerical studies in Burdett et. al. (2017) and Burdett and Kozan (2016) which were "over-detailed" in that respect.

### 3.2. Techniques

In this section the quantitative techniques behind HOPLITE are described. The main procedures facilitated by HOPLITE are summarised in Figure 2. Table 2 then describes all notation, parameters, and terminology needed to understand them. Decision variables are labelled DVAR in Table 2.

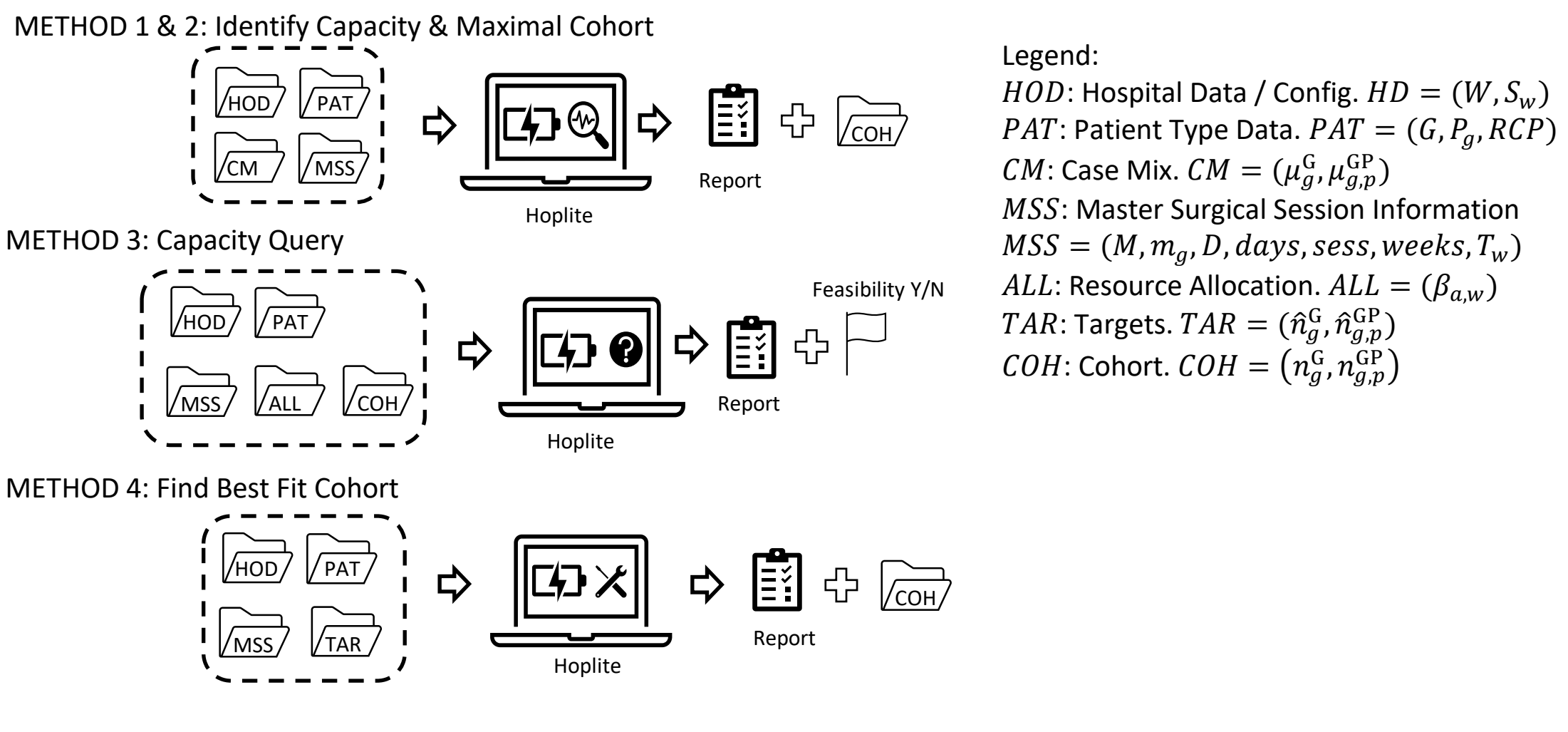

**Figure 2.** Primary quantitative methods of HOPLITE

**Table 2.** Notation and key parameters.

| SYMBOL | MEANING | ADDITIONAL NOTES |
|---|---|---|
| $\boldsymbol{G}$ | Set of patient types | |
| $\boldsymbol{P_g}$ | Set of patient sub types within group $g$ | |
| $\boldsymbol{A}$ | Set of all surgical/medical activities | |
| $\boldsymbol{A_{g,p}}$ | The set of activities for patients of sub type $(g,p)$ | • $A_{g,p} = \{(g,p,1),(g,p,2),(g,p,3)\}$ |
| $\boldsymbol{K_{g,p}}$ | The number of activities in the profile for patient sub type $(g,p)$. | • $K_{g,p} = 3$ |
| $\boldsymbol{t_{(g,p,k)}}$ | Time (deterministic) to perform activity type $(g,p,k)$. Otherwise viewed as the resource time consumed. Unit is hours. | • $k=1 \Rightarrow$ theatre "surgery" time<br>• $k=2 \Rightarrow$ ward postop time<br>$k=3 \Rightarrow$ intensive care time |
| $\boldsymbol{A_w}$ | The set of activities performed in area $w$ | |
| $\boldsymbol{OT, ICU}$ | The set of theatres and intensive care beds | |
| $\boldsymbol{WARDS}$ | The set of recovery wards | |
| $\boldsymbol{W}$ | Set of hospital areas. $W = OT \cup ICU \cup WARDS$ | • Areas include wards, theatres, and the intensive care area |
| $\boldsymbol{S_w}$ | Number of treatment spaces in area $w$. | • Typically refers to number of beds present. |
| $\boldsymbol{Beds}$ | Total number of ward recovery beds | • $Beds = \sum_{w \in WARDS} S_w$ |
| $\boldsymbol{ICBeds}$ | Total number of intensive care beds | |
| $\boldsymbol{Use_w}$ | Time usage of area $w$ | • Unit is hours |
| $\boldsymbol{W_a}$ | The set of areas that activity $a$ can be performed in. This set describes alternative options. | • There must be at least one option, i.e., $\lvert W_a \rvert \geq 1$ |
| $\boldsymbol{T_w}$ | The time availability of hospital area $w \in W$ | • Unit is hours |
| $\boldsymbol{\mu_g^{\mathrm{G}}}$ | Case mix. The proportion of patients of type $g$ amongst the entire patient cohort. | • $\sum_g \mu_g^{\mathrm{G}} = 1$<br>• $0 \leq \mu_g^{\mathrm{G}} \leq 1$ |
| $\boldsymbol{\mu_{g,p}^{\mathrm{GP}}}$ | Sub type mix. The proportion of patients of sub type $p$ amongst type $g$ | • $\sum_{p \in P_g} \mu_{g,p}^{\mathrm{GP}} = 1 \ \forall g \in G$<br>• $0 \leq \mu_{g,p}^{\mathrm{GP}} \leq 1$ |
| $\boldsymbol{\hat{n}_g^{\mathrm{G}}, \hat{n}_{g,p}^{\mathrm{GP}}}$ | Target case mix and sub mix | • Conditionally, $\hat{n}_g^{\mathrm{G}} = \mu_g^{\mathrm{G}} \widehat{\mathbb{N}}$<br>Conditionally, $\hat{n}_{g,p}^{\mathrm{GP}} = \mu_{g,p}^{\mathrm{GP}} \hat{n}_g^{\mathrm{G}}$ |
| $\widehat{\mathbb{N}}$ | Target number of patients | • $\widehat{\mathbb{N}} = \sum_g \hat{n}_g^{\mathrm{G}}$ or $\widehat{\mathbb{N}} = \sum_g \sum_{p \in P_g} \hat{n}_{g,p}^{\mathrm{GP}}$ |
| $\boldsymbol{\hat{\beta}_{a,w}}$ | Target allocation | • Consequence: $\hat{n}_{g,p}^{\mathrm{GP}} = \sum_{w \in W_a} \hat{\beta}_{a,w}$ |
| $\$_{g,p}^{\mathrm{GP}}$ | The revenue specified for a specific sub type | • $\$_g^{\mathrm{G}} = \sum_{p \in P_g} n_{g,p}^{\mathrm{GP}} \$_{g,p}^{\mathrm{GP}}$ and $\hat{\$} = \sum_g \$_g^{\mathrm{G}}$ |
| $\mathbb{N}$ | Total number of patients to be treated [DVAR] | • $\mathbb{N} = \sum_g n_g^{\mathrm{G}} = \sum_g \sum_{p \in P_g} n_{g,p}^{\mathrm{GP}}$ |
| $\boldsymbol{n_g^{\mathrm{G}}}$ | Number of type $g \in G$ patients to be treated [DVAR]. The name/descriptor of the type is denoted $\mathbb{ID}_g$. | • $G$ is the set of patient types<br>• $n_g^{\mathrm{G}} = \sum_{p \in P_g} n_{g,p}^{\mathrm{GP}} \quad \forall g \in G$<br>• Conditionally, $n_g^{\mathrm{G}} = \mu_g^{\mathrm{G}} \mathbb{N}$ |
| $\boldsymbol{n_{g,p}^{\mathrm{GP}}}$ | Number of sub type $p \in P_g$ patients to be treated [DVAR]. The name/descriptor of the sub type is denoted $\mathbb{ID}_{g,p}$. | • $P_g$ is the set of patient sub types.<br>• Conditionally, $n_{g,p}^{\mathrm{GP}} = \mu_{g,p}^{\mathrm{GP}} n_g^{\mathrm{G}}$ |
| $\boldsymbol{\beta_{a,w}}$ | The number of activities of type $a = (g,p,k)$ assigned to hospital area $w$, where $g$ is the patient type, $p$ is the sub type, and $k$ is the activity type indexer [DVAR]. | • $1 \leq k \leq K_{g,p}$<br>• $\beta_{a,w} = 0 \quad \forall a \in A, \forall w \in W \backslash W_a$ |
| $\boldsymbol{M}$ | Total number of sessions in the MSS | • $M = weeks \times days \times sess \times \lvert OTU \rvert$<br>$M \times D$ is the total time availability for surgical activities |
| $\boldsymbol{D}$ | Duration of sessions in the MSS | • Typically, $D = 4$ hrs |
| $\boldsymbol{days}$ | Days per week in that theatres are running in MSS | • Typically, $days = 5$ |
| $\boldsymbol{sess}$ | Sessions per day in MSS | • Typically, $sess = 2$ |
| $\boldsymbol{weeks}$ | Number of weeks over which an assessment is performed | • $weeks \geq 1$ |
| $\boldsymbol{m_g}$ | Number of sessions assigned to patients of type $g$. Real valued and may take fractional values | • $m_g = \mu_g^{\mathrm{G}} M$ or else it is chosen such that $m_g \leq \mu_g^{\mathrm{G}} M$<br>• $\sum_g m_g = M$ |

### 3.2.1. Basic Assessment of Capacity

An approach to perform a basic assessment of capacity is first described. This approach involves static calculations and does not require the application of a mathematical optimization model. The main idea behind the approach is the restriction of capacity by a particular resource type. Evaluating changes to key parameters like $M, D, days, sess, weeks, S_w$ can be quickly evaluated and this permits hospital capacity expansion planning and other what-if assessments. Resource utilisation information can be computed, viewed, and actioned.

**Method 1.** Given an MSS template and a pre-established allocation of theatre sessions to specialties (i.e., $m_g$) an important task is to identify the patients, denoted $\mathbb{N}, n_g^{\mathrm{G}}, n_{g,p}^{\mathrm{GP}}$, that can be theoretically treated. If we assume that there is only one ward option for each postop activity, the static calculations described by equation (1) and (3) are sufficient to determine the case mix and sub mix. Only theatre usage is restricted with this approach, and all other resource type usage is unrestricted.

$$n_g^{\mathrm{G}} = \frac{m_g D}{\sum_{p \in P_g} \mu_{g,p}^2 t_{g,p,1}} \ \forall g \in G \tag{1}$$

Term, $m_g D$ is the total time available, and the denominator is the weighted average surgery time.

**Method 2.** A second query of a similar nature can also be performed using Equation (2). With this approach it is assumed that output is restricted by ward bed availability, and not theatre availability. Beds are assumed available 24 hrs/day and 7 days/week; hence availability is 168 hrs/week per bed. Parameter $ward_{g,p}$ is the specific ward assigned to perform the postop activity designated $(g, p, 2)$.

$$n_g^{\mathrm{G}} = (168 \times weeks \times S_w) / \sum_{p \in P_g} \mu_{g,p}^{\mathrm{GP}} \left(t_{g,p,1} + t_{g,p,2}\right) \ \text{where } w = ward_{g,p} \tag{2}$$

In equation (2), the denominator represents the weighted average time utilisation in a ward bed. In many hospitals, ward beds are acquired and made vacant before surgery is performed. Hence, the total time consumed is $t_{g,p,1} + t_{g,p,2}$.

For both methods, the following calculations are made:

$$n_{g,p}^{\mathrm{GP}} = \mu_{g,p}^{\mathrm{GP}} n_g^{\mathrm{G}} \ \forall g \in G, \forall p \in P_g, \ \mathbb{N} = \sum_{g \in G} n_g^{\mathrm{G}} \tag{3}$$

$$BedUse = \sum_{g \in G} \sum_{p \in P_g} n_{g,p}^{\mathrm{GP}} \left(t_{g,p,1} + t_{g,p,2}\right), BedUt = 100 \times BedUse/(168 \times Beds) \tag{4}$$

$$BedUse_w = \sum_{g \in G} \sum_{p \in P_g | ward_{g,p} = w} n_{g,p}^{\mathrm{GP}} \left(t_{g,p,1} + t_{g,p,2}\right) \ \forall w \in WARDS,$$
$$BedUt_w = 100 \times BedUse_w/(168 \times S_w) \tag{5}$$

$$OtUse = \sum_{g \in G} \sum_{p \in P_g} n_{g,p}^{\mathrm{GP}} \left(t_{g,p,1}\right), OtUt = 100 \times OtUse/(M \times D) \tag{6}$$

$$IcUse = \sum_{g \in G} \sum_{p \in P_g} n_{g,p}^{\mathrm{GP}} \left(t_{g,p,3}\right), IcUt = 100 \times IcUse/(168 \times IcBed) \tag{7}$$

### 3.2.2. Advanced Assessments of Capacity

The methods in this section constitute a more advanced assessment of capacity. Identifying the maximum number of patients treatable, over time, given specified time availabilities of ward beds, theatres, and intensive care beds, and some notion of case mix, provides useful information to hospital managers and planners. A mathematical model must be applied to determine the exact number of each type/sub-type, namely $n_g^{\mathrm{G}}$ and $n_{g,p}^{\mathrm{GP}}$, and a resource allocation $\beta_{a,w}$, because of the competition for common resources, and the availability of optional locations and resources to choose from. It is not possible to know how to assign resources just by looking, or to apply static calculations. Conceptually this decision problem is similar to multi-knapsack and bin-packing problems.

**Method 1.** When a case mix is viewed as the proportion of all patients treated, the necessary optimization model is as follows:

Maximize $\mathbb{N}$
Subject to

$$n_{g,p}^{\mathrm{GP}} = \sum_{w \in W_a} \beta_{a,w} \quad \forall a \in A \tag{8}$$

$$\sum_{\forall a \in A_w} \beta_{a,w} t_a \le T_w S_w \ \forall w \in W \tag{9}$$

- $$\sum_{g \in G} \sum_{p \in P_g} \beta_{(g,p,1),OT} t_{g,p,1} \le T_{OTU} S_{OTU} \tag{10}$$
- $$\sum_{g \in G} \sum_{p \in P_g} \beta_{(g,p,2),w} \left(t_{g,p,1} + t_{g,p,2}\right) \le T_w S_w \ \forall w \in WARD \tag{11}$$
- $$\sum_{g \in G} \sum_{p \in P_g} \beta_{(g,p,3),ICU} t_{g,p,3} \le T_{ICU} S_{ICU} \tag{12}$$

$$n_{g,p}^{\mathrm{GP}} \ge u_{g,p}^{\mathrm{GP}} n_g^{\mathrm{G}} \ \forall g \in G, \forall p \in P_g \tag{13}$$

$$n_g^{\mathrm{G}} \ge \mu_g^{\mathrm{G}} \mathbb{N} \ \forall g \in G \tag{14}$$

$$\beta_{a,w} \ge 0 \quad \forall a \in A, \forall w \in W_a \tag{15}$$

Equation (8) is a balance equation, providing a link between the allocations made and the number of patients of a given sub type. Equation (9) is a generic constraint for the time availability restriction of hospital area. Equation (10)-(12) are respectively specific to OT, WARD, and ICU areas. Typically, $T_w = weeks \times 7 \times 24 \quad \forall w \in W \backslash OTU$ and $T_{OTU} = weeks \times days \times sess \times D$. Equation (13) and (14) are "administrative" constraints enforcing specified case mix and sub mix proportions.

**Method 2.** Anecdotally, we have observed that hospitals do not always view the case mix as a proportion of the whole. Instead, they view case mix relative to the theatre time allocated to each group type. In that scenario a different optimization model is required:

Maximize $\sum_g n_g^{\mathrm{G}}$
Subject to

Constraint (8), (9) or (10) – (12), (13), (15)

$$\sum_{p \in P_g} n_{g,p}^{\mathrm{GP}} t_{(g,p,1)} \le m_g D \ \forall g \in G \tag{16}$$

$$n_g^{\mathrm{G}} \ge 0 \ \forall g \in G \tag{17}$$

Constraint (16) is added to restrict output by session time allocated. The number of sessions assigned to each group is $m_g = \mu_g^{\mathrm{G}} M$. To simplify matters, it is recommended that all theatres are aggregated into a single area within the model input data. Hence, the number of spaces in that area is the number of theatres. It can be assumed that there is only one ICU in most hospitals, and all intensive care beds are there. If that is not the case, it is recommended to aggregate all the ICU beds into one area too.

### 3.2.3. Evaluating Case Mix Feasibility and Requirements

Testing the feasibility of a selected case mix and/or resource allocation is another useful capacity query. All intensions specified by a planner are hereby designated as targets and denoted using the variables $\hat{n}_g^{\mathrm{G}}, \hat{n}_{g,p}^{\mathrm{GP}}, \hat{\beta}_{a,w}$. If only $\hat{\beta}_{a,w}$ is defined, then the usage level of each resource can be directly evaluated using equation (4)-(7). If no resource is over-used, then feasibility of the allocation is verified. In these circumstances, $n_{g,p}^{\mathrm{GP}} = \sum_{w \in W_a} \hat{\beta}_{a,w}$ and $n_g^{\mathrm{G}} = \sum_{p \in P_g} n_{g,p}^{\mathrm{GP}}$. If $\hat{n}_g^{\mathrm{G}}$ or $\hat{n}_{g,p}^{\mathrm{GP}}$ or both are defined, but $\hat{\beta}_{a,w}$ is not, it is necessary to solve the model from Section 3.2.2 to identify whether a feasible resource allocation can be obtained. Two additional constraints are, however added, namely $n_g^{\mathrm{G}} = \hat{n}_g^{\mathrm{G}} \ \forall g \in G$ and $n_{g,p}^{\mathrm{GP}} = \hat{n}_{g,p}^{\mathrm{GP}} \ \forall g \in G, \forall p \in P_g$.

Determining the number of surgical sessions required in the MSS for each specialty is another useful piece of information. Given target $\hat{n}_g^{\mathrm{G}}$ then $m_g = \frac{\hat{n}_g^{\mathrm{G}} \sum_{p \in P_g} \mu_{g,p}^{\mathrm{GP}} t_{g,p,1}}{D}$. The numerator is the weighted average multiplied by the number of patients. Similarly, given $\hat{n}^{\mathrm{GP}}_{g,p}$, then $m_g = \frac{\sum_{p\in P_g} \hat{n}^{\mathrm{GP}}_{g,p} t_{g,p,1}}{D}$ $\forall g \in G$.

### 3.2.4. Identifying a Case Mix that Best Meets Given Targets

In Section 3.2.3, the feasibility of a target case and/or sub mix was discussed. If those targets are infeasible, a pertinent question is, what is the closest feasible case mix to the designated targets. The model shown in Section 3.2.2 should be solved such that the deviation between the targets and the actual values is minimized. If the targets can be met exactly, then the objective value will be zero.

For this capacity query it is assumed that exceeding one target is not sufficient to counterbalance any unmet target. In other words, the constraints $n^{\mathrm{G}}_g \le \hat{n}^{\mathrm{G}}_g$ $\forall g$ and or $n^{\mathrm{GP}}_{g,p} \le \hat{n}^{\mathrm{GP}}_{g,p}$ $\forall g, \forall p \in P_g$ should be added. The following upper bound is also needed:

$$\beta_{(g,p,2),w} \le \frac{T_w S_w}{(t_{g,p,1}+t_{g,p,2})} \quad \forall (g,p,2) \in A|, \forall w \in W_{(g,p,2)} \tag{18}$$

This constraint is helpful when solving the decision model with a non-linear solver. Adding upper bounds is general advice provided on the OpenSolver website. As theatres are deemed generic, and there is one ICU, the following upper bounds could also be imposed:

$$\beta_{(g,p,1),w} \le \frac{T_{OTU} S_{OTU}}{(t_{g,p,1})} \quad \forall (g,p,1) \in A, \forall w \in W_{(g,p,1)} \tag{19}$$

$$\beta_{(g,p,3),w} \le \frac{T_{ICU} S_{ICU}}{(t_{g,p,3})} \quad \forall (g,p,3) \in A, \forall w \in W_{(g,p,3)} \tag{20}$$

Several variant objective functions may be used, and these relate to different “targeting options”.

**Target Option 1 (TO1)**: Given targets $\hat{n}^{\mathrm{G}}_g$ and importance weights $\omega_g$, the objective is either of the following:

$$\text{Minimize } \left\| n^{\mathrm{G}} - \hat{n}^{\mathrm{G}} \right\|_1 = \textstyle\sum_g \omega_g \left| n^{\mathrm{G}}_g - \hat{n}^{\mathrm{G}}_g \right| \equiv \sum_g \omega_g \left( \hat{n}^{\mathrm{G}}_g - n^{\mathrm{G}}_g \right) \tag{21}$$

$$\text{Minimize } \left\| n^{\mathrm{G}} - \hat{n}^{\mathrm{G}} \right\|_2 = \left( \textstyle\sum_g \omega_g \left( n^{\mathrm{G}}_g - \hat{n}^{\mathrm{G}}_g \right)^2 \right)^{1/2} \text{ or Minimize } \sum_g \omega_g \left( n^{\mathrm{G}}_g - \hat{n}^{\mathrm{G}}_g \right)^2 \tag{22}$$

The absolute value term $\left| n^{\mathrm{G}}_g - \hat{n}^{\mathrm{G}}_g \right|$ can be replaced with $(\hat{n}^{\mathrm{G}}_g - n^{\mathrm{G}}_g)$ in equation (21) as $n^{\mathrm{G}}_g \le \hat{n}^{\mathrm{G}}_g$. This makes equation (21) a linear expression. Equation (22), however, is non-linear and that necessitates the application of a non-linear solver. Minimizing the sum of squares $\sum_g \omega_g \left( n^{\mathrm{G}}_g - \hat{n}^{\mathrm{G}}_g \right)^2$ is a viable alternative. That expression is a multivariate quadratic function (see Appendix A-1). As such, it is hypothesised that the decision model may be solved more efficiently via Quadratic Programming techniques, rather than generic non-linear solvers.

If sub-types are specified, then targets $\hat{n}^{\mathrm{GP}}_{g,p}$ may be defined as follows, $\hat{n}^{\mathrm{GP}}_{g,p} = \mu^{\mathrm{GP}}_{g,p} \hat{n}^{\mathrm{G}}_g$. These can also be chosen to be explicitly met too.

**Target Option 2 (TO2)**: Given targets $\hat{n}^{\mathrm{GP}}_{g,p}$ the objective is either of the following:

$$\text{Minimize } \left\| n^{\mathrm{GP}} - \hat{n}^{\mathrm{GP}} \right\|_{1,1} = \textstyle\sum_g \omega_g \sum_{p\in P_g} \left( \hat{n}^{\mathrm{GP}}_{g,p} - n^{\mathrm{GP}}_{g,p} \right) \text{ or } \sum_g \omega_g \sum_{p\in P_g} \left( \hat{n}^{\mathrm{GP}}_{g,p} - n^{\mathrm{GP}}_{g,p} \right) / \hat{n}^{\mathrm{GP}}_{g,p} \tag{23}$$

$$\text{Minimize } \left\| n^{\mathrm{GP}} - \hat{n}^{\mathrm{GP}} \right\|_{2,2} = \left( \textstyle\sum_g \omega_g \sum_{p\in P_g} \left( n^{\mathrm{GP}}_{g,p} - \hat{n}^{\mathrm{GP}}_{g,p} \right)^2 \right)^{\frac{1}{2}} \tag{24}$$

Like equation (21), equation (23) is linear, because $n^{\mathrm{GP}}_{g,p} \le \hat{n}^{\mathrm{GP}}_{g,p}$. Equation (24), however, is non-linear like (22). It is also a quadratic function (see Appendix A-2).

**Target Option 3 (TO3)**: Given both targets $\hat{n}_g^{\mathrm{G}}$ and $\hat{n}_{g,p}^{\mathrm{GP}}$ the objective is $\|n^{\mathrm{GP}} - \hat{n}^{\mathrm{GP}}\| + \|n^{\mathrm{G}} - \hat{n}^{\mathrm{G}}\|$. If $\hat{n}_g^{\mathrm{G}} < \sum_{p \in P_g} \hat{n}_{g,p}^{\mathrm{GP}}$ then the definition of $\hat{n}_g^{\mathrm{G}}$ is inconsistent, and it makes sense to redefine $\hat{n}_g^{\mathrm{G}} = \sum_{p \in P_g} \hat{n}_{g,p}^{\mathrm{GP}}$ automatically. In general, the update $\hat{n}_g^{\mathrm{G}} = \max\left(\hat{n}_g^{\mathrm{G}}, \sum_{p \in P_g} \hat{n}_{g,p}^{\mathrm{GP}}\right)$ is appropriate.

**Further things to note.** The alternative norms shown do different things and will not produce the same optimal decisions. The 2-norm will ensure most of the targets are met or are minimally unmet. This is because any difference is squared. The 1-norm can lead to solutions which meet some targets exactly at the expense of others. If the objective is zero, i.e., the norms are zero, then there may be a better solution that satisfies all targets, but with higher overall throughput. A better solution can for instance be obtained by setting the objective to Maximize $\mathbb{N}$, Maximize $\|n^{\mathrm{G}} - \hat{n}^{\mathrm{G}}\|$, or Maximize $\|n^{\mathrm{GP}} - \hat{n}^{\mathrm{GP}}\|$, and by adding the constraints $n_g^{\mathrm{G}} \geq \hat{n}_g^{\mathrm{G}}$ and $n_{g,p}^{\mathrm{GP}} \geq \hat{n}_{g,p}^{\mathrm{GP}}$.

The previously defined objectives penalise all differences that occur. Exceeding targets is desirable, but exceeding a target is not deemed sufficient to counterbalance any unmet target. Hence, the following alternatives are considered invalid.

Minimize $Deficits = \sum_g \delta_g^{\mathrm{G}}$ where $\delta_g^{\mathrm{G}} = \max\,(\hat{n}_g^{\mathrm{G}} - n_g^{\mathrm{G}}, 0)$ (25)

Minimize $Deficits = \sum_g \sum_p \delta_{g,p}^{\mathrm{GP}}$ where $\delta_{g,p}^{\mathrm{GP}} = \max\,(\hat{n}_{g,p}^{\mathrm{GP}} - n_{g,p}^{\mathrm{GP}}, 0)$ (26)

Minimize $Deficits = \max\,(\hat{\mathbb{N}} - \mathbb{N}, 0)$ (27)

The target capacity is computed as $\hat{\mathbb{N}} = \sum_g \hat{n}_g^{\mathrm{G}}$ or as $\hat{\mathbb{N}} = \sum_g \sum_{p \in P_g} \hat{n}_{g,p}^{\mathrm{GP}}$. Minimizing $\|\mathbb{N} - \hat{\mathbb{N}}\|$ as the objective is another option.

When $\hat{n}_{g,p}^{\mathrm{GP}}$ is specified, it would be tempting to conclude that $\hat{n}_g^{\mathrm{G}} = \sum_{p \in P_g} \hat{n}_{g,p}^{\mathrm{GP}}$ as well. However, minimising $\|n^{\mathrm{G}} - \hat{n}^{\mathrm{G}}\|$ will not guarantee that $\|n^{\mathrm{GP}} - \hat{n}^{\mathrm{GP}}\|$ is minimized. Also, $\|n^{\mathrm{G}} - \hat{n}^{\mathrm{G}}\| \neq \|n^{\mathrm{GP}} - \hat{n}^{\mathrm{GP}}\|$ except when the 1-norm is applied.

**Lemma 1**: $\|n^{\mathrm{G}} - \hat{n}^{\mathrm{G}}\|_1 = \|n^{\mathrm{GP}} - \hat{n}^{\mathrm{GP}}\|_{1,1}$ where $\|A\|_{x,x} = \|\mathrm{vec}(A)\|_x = \sum_i \sum_j a_{i,j}$. **Proof**. By definition, $\|n^{\mathrm{G}} - \hat{n}^{\mathrm{G}}\|_1 \equiv \sum_g (\hat{n}_g^{\mathrm{G}} - n_g^{\mathrm{G}})$. As $n_g^{\mathrm{G}} = \sum_{p \in P_g} n_{g,p}^{\mathrm{GP}}$ and $\hat{n}_g^{\mathrm{G}} = \sum_{p \in P_g} \hat{n}_{g,p}^{\mathrm{GP}}$, then $\|n^{\mathrm{G}} - \hat{n}^{\mathrm{G}}\|_1 = \sum_g \left(\sum_{p \in P_g} \hat{n}_{g,p}^{\mathrm{GP}} - \sum_{p \in P_g} n_{g,p}^{\mathrm{GP}}\right) = \sum_g \sum_{p \in P_g} (\hat{n}_{g,p}^{\mathrm{GP}} - n_{g,p}^{\mathrm{GP}}) = \|n^{\mathrm{GP}} - \hat{n}^{\mathrm{GP}}\|_{1,1}$.

If the two norm is applied (i.e., see equation (22) and (24)), then $(n_g^{\mathrm{G}} - \hat{n}_g^{\mathrm{G}})^2 \neq \sum_{p \in P_g} (n_{g,p}^{\mathrm{GP}} - \hat{n}_{g,p}^{\mathrm{GP}})^2$ and Lemma 2 holds true.

**Lemma 2**: $\|n^{\mathrm{GP}} - \hat{n}^{\mathrm{GP}}\|_{2,2} \leq \|n^{\mathrm{G}} - \hat{n}^{\mathrm{G}}\|_2$. **Proof**. By Definition, $\left(\|n^{\mathrm{G}} - \hat{n}^{\mathrm{G}}\|_2\right)^2 = \sum_g (n_g^{\mathrm{G}} - \hat{n}_g^{\mathrm{G}})^2$ and

$\left(\|n^{\mathrm{GP}} - \hat{n}^{\mathrm{GP}}\|_{2,2}\right)^2 = \left(\|\mathrm{vec}(n^{\mathrm{GP}} - \hat{n}^{\mathrm{GP}})\|_2\right)^2 = \sum_g \sum_{p \in P_g} (n_{g,p}^{\mathrm{GP}} - \hat{n}_{g,p}^{\mathrm{GP}})^2$ . Given that $n_{g,p}^{\mathrm{GP}} = \mu_{g,p}^{\mathrm{GP}} n_g^{\mathrm{G}}$, then:

$\left(\|n^{\mathrm{GP}} - \hat{n}^{\mathrm{GP}}\|_{2,2}\right)^2 = \sum_g \sum_{p \in P_g} (\mu_{g,p}^{\mathrm{GP}} n_g^{\mathrm{G}} - \mu_{g,p}^{\mathrm{GP}} \hat{n}_g^{\mathrm{G}})^2 = \sum_g (n_g^{\mathrm{G}} - \hat{n}_g^{\mathrm{G}})^2 \sum_{p \in P_g} (\mu_{g,p}^{\mathrm{GP}})^2$. As $\sum_{p \in P_g} (\mu_{g,p}^{\mathrm{GP}})^2 \leq 1$, then $\left(\|n^{\mathrm{GP}} - \hat{n}^{\mathrm{GP}}\|_{2,2}\right)^2 \leq \left(\|n^{\mathrm{G}} - \hat{n}^{\mathrm{G}}\|_2\right)^2$.

## 4. HOPLITE PDST

In this section our PDST is introduced, and the main capabilities and features are demonstrated.

### 4.1. Software Organisation and Design.

The PDST provides several graphical interfaces. After loading the PDST, the main application window is shown. This leads users to the main options window, see Figure 3. From this window, the user can then choose which decision making, evaluation or assessment task they would like to do. Once completed, users are returned to the main window. Decision-making tasks require the solution of a mathematical optimization model, but evaluation tasks do not. Model generation is automated using VBA and solution is via the application of the Excel Solver or OpenSolver. OpenSolver extracts the optimization model defined in the spreadsheet, which is then written to a file and passed to the Coin - CBC optimization engine to solve. The result is then read in, and automatically loaded back into the spreadsheet. When using OpenSolver, a decision model can be generated by applying the OpenSolvers' inbuilt commands and functions. These are different to Solvers'. However, OpenSolver also takes as input a model generated using Solver commands and functions.

Each task is performed in a different window at present. Some tasks could be aggregated into one window, however, that would result in a more congested and cluttered appearance.

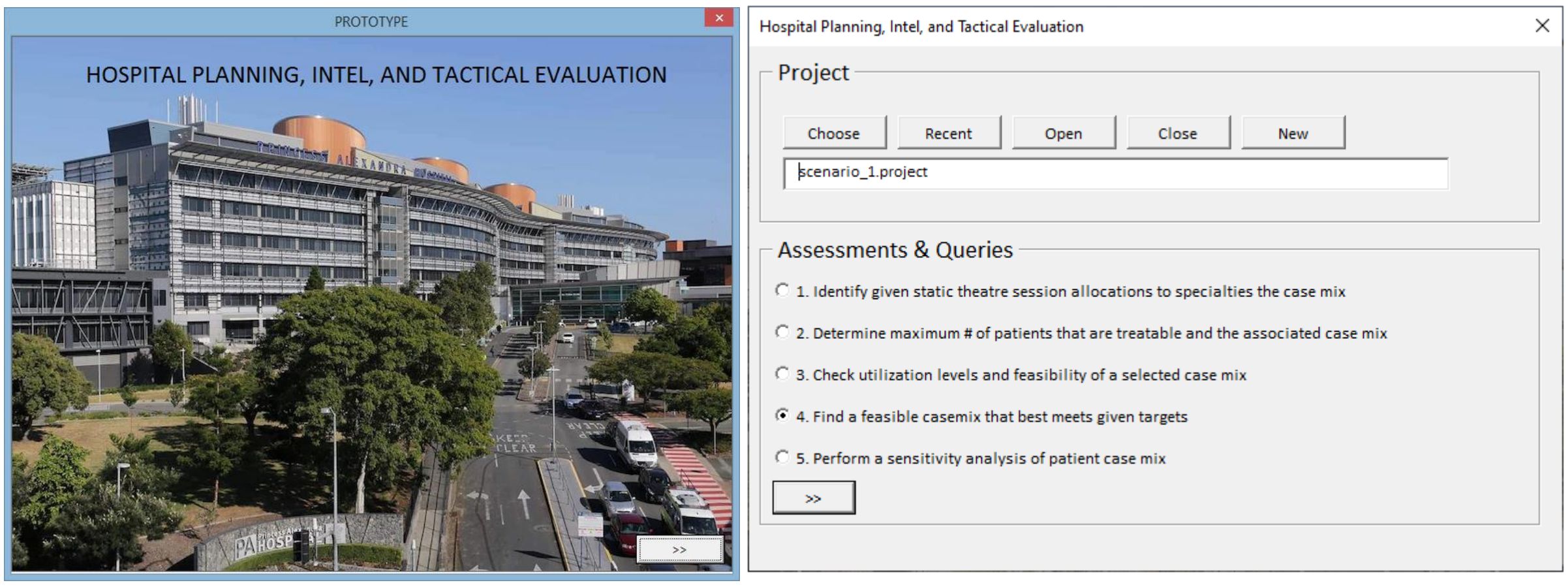

**Figure 3.** Main application and options window

To select an existing project to load, the "Choose" or "Recent" button is first pressed. A project however is not loaded until "Open" is pressed. Once a valid project is loaded, the options are enabled; prior to this, they are disabled. A new project can be set up either using HOPLITE or externally. The "New" button brings up the window shown in Figure 4. A plain example is shown in that figure for ease of understanding. Frame 1 asks for a project name first. When the command button ">>" in that frame is pressed, a folder is created, and necessary data files are created. These will be discussed in Section 4.2. In Frame 2, all theatres, ward beds, and intensive care beds can be defined. The ListView in that frame only shows the wards, as these are deemed unique objects. In contrast, theatres and intensive beds are not. In this PDST there is assumed one operating theatre unit to encapsulate all the theatres and one intensive care unit to encapsulate all the intensive care beds. Pressing the command button ">>" in that frame permits the user to move on to Frame 3 to designate patient types and sub types. The time required for surgery, intensive care and postop activities must be input for each sub type. These values constitute the resource consumption profile discussed at the start of Section 3. In theory these values could be populated programmatically for larger datasets and input directly from historical records stored in a data warehouse.

To add a patient sub type, it is first necessary to select a row from ListView 2. This requirement explicitly links a patient type to a patient sub type. To add a ward option for each patient sub type, it is necessary to select a row from ListView 3 and to press the "+" button to the right of the "Ward Option" combo box. This combo box is populated by the information input in Frame 2. Pressing the command button ">>" at the bottom of Frame 3 finalises the project by extracting all information on the UserForm and writing it to file in the proper data format.

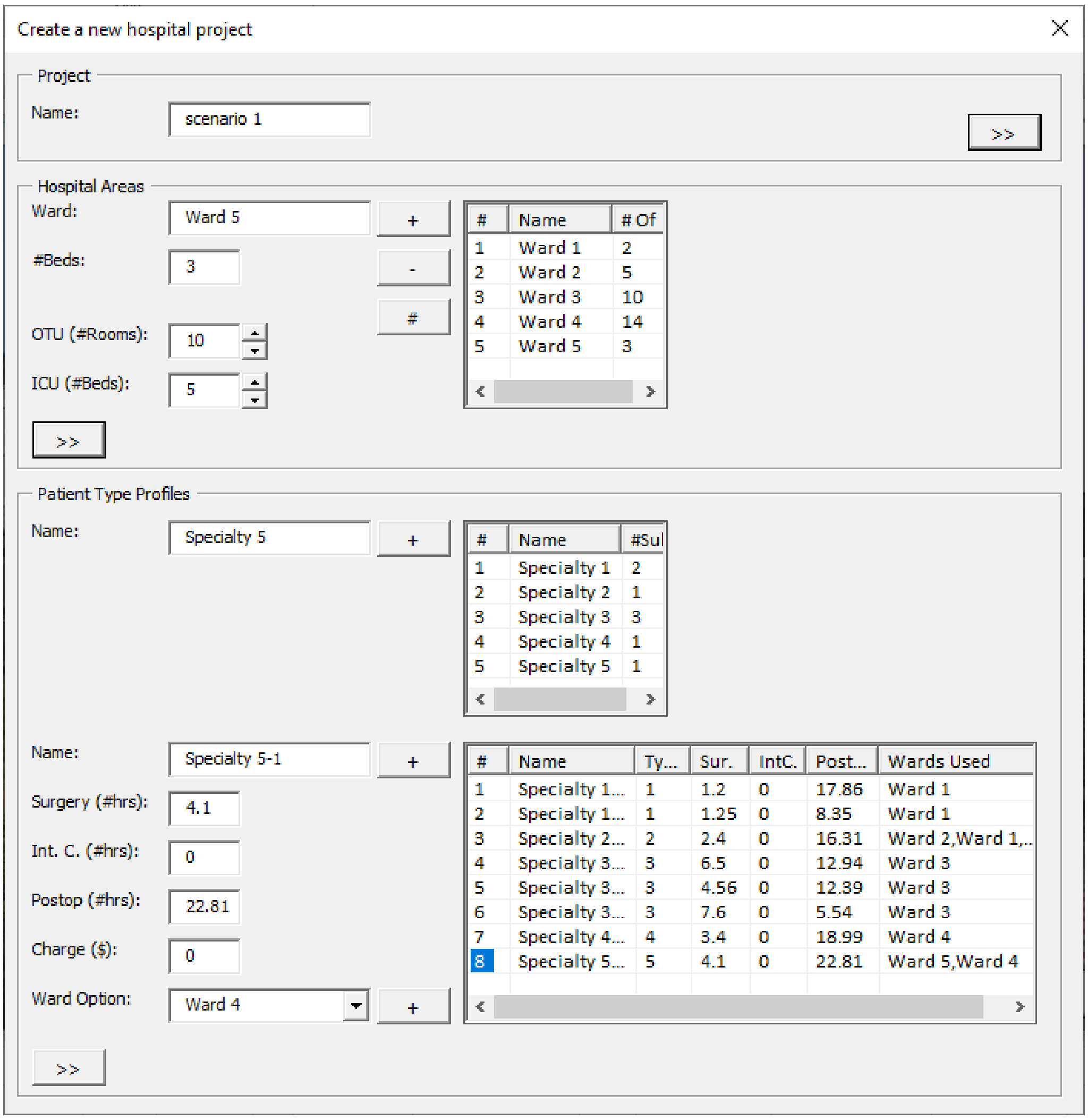

**Figure 4.** Project creation window

### 4.1. Data Requirements and Inputs

HOPLITE requires various information, and this is input from a variety of primary and secondary (a.k.a., auxiliary) files. In this section details of those files are summarised in Table 3 and 4. The set bracketing notation "{(...)}" implies multiple line inputs of tuples (...).

The first requirement is a project file that specifies the primary inputs, namely the hospital configuration and the patient type information. In the configuration file, the hospital areas (a.k.a., resources) are input. Given our development assumptions, it is not necessary to differentiate between different theatres and intensive care beds. However, it is necessary to differentiate between different wards. This is strictly necessary for the quantitative techniques and for reporting. As such, each ward has an identifier $\mathbb{ID}_w$.

**Table 3.** Main project file and primary inputs

| File Type: *.project | File Type: *.config | File Type: *.patient |
|---|---|---|
| Project Name,name | Intensive Care Beds, $S_{ICU}$ | Patient Types,$|G|$ |
| Hospital Configuration,*.config | Theatres, $S_{OTU}$ | Patient Type, |
| Patient Information,*.patient | Wards,$|WARD|$ | $\{[g], \mathbb{ID}_g, |P_g|\}$ |
| Case Mix,*.mix | Ward Information, | Patient Sub Type, |
| Session,*.session | $\{([w], \mathbb{ID}_w, S_w)\}$ | $\{[g][p], \mathbb{ID}_{g,p}\}$ |
| Targets,*.target | | Profile, |
| Allocation,*.alloc | | $\{[g][p], t_{g,p,1}, t_{g,p,2}, t_{g,p,3}, A_{g,p}\}$ |
| | | Revenue, |
| | | $\{[g][p], \$^2_{g,p}\}$ |

A concrete example is shown below for a small fictional hospital. There are a total of 34 beds placed across five wards. Of the different patient sub types, two have optional ward locations. In total there are eight patient profiles.

**Table 4.** Concrete examples of primary files shown in Table 3

| scenario_1.project | scenario_1.config | scenario_1.patient |
|---|---|---|
| Project Name,scenario_1<br>Hospital Configuration,scenario_1.config<br>Patient Information,scenario_1.patient<br>Case Mix,<br>Session,<br>Targets,<br>Allocation, | Intensive Care Beds,5<br>Theatres,10<br>Wards,5<br>Ward Info,<br>[1],Ward 1,2<br>[2],Ward 2,5<br>[3],Ward 3,10<br>[4],Ward 4,14<br>[5],Ward 5,3 | Patient Types,5<br>Patient Type,<br>[1],Specialty 1,2<br>[2],Specialty 2,1<br>[3],Specialty 3,3<br>[4],Specialty 4,1<br>[5],Specialty 5,1<br>Patient Sub Type,<br>[1][1],Specialty 1-1<br>[1][2],Specialty 1-2<br>[2][1],Specialty 2-1<br>[3][1],Specialty 3-1<br>[3][2],Specialty 3-2<br>[3][3],Specialty 3-3<br>[4][1],Specialty 4-1<br>[5][1],Specialty 5-1<br>Profile,<br>[1][1],0,1.2,17.86,Ward 1<br>[1][2],6,1.25,8.35,Ward 1<br>[2][1],0,2.4,16.31,Ward 2,Ward 1,Ward 5<br>[3][1],0,6.5,12.94,Ward 3<br>[3][2],0,4.56,12.39,Ward 3<br>[3][3],0,7.6,5.54,Ward 3<br>[4][1],0,3.4,18.99,Ward 4<br>[5][1],12,4.1,22.81,Ward 5,Ward 4<br>Revenue,<br>[1][1],1000.0<br>[1][2],1500.0<br>[2][1],600.0<br>[3][1],2500.0<br>[3][2],6000.0<br>[3][3],3700.0<br>[4][1],10000.0<br>[5][1],5500.0 |

Other inputs include a case mix, sub mix, session allocation, targets, and resource allocation. These are loaded from the following files:

**Table 5.** Auxiliary input files

| File Type: *.mix | File Type: *.session | File Type: *.alloc | File Type: *.target |
|---|---|---|---|
| Case Mix,<br>$\{([g], \mathbb{ID}_g, \mu_g^{\mathrm{G}})\}$<br>Sub Mix,<br>$\{([g][p], \mathbb{ID}_{g,p}, \mu_{g,p}^{\mathrm{GP}})\}$ | Patient type,<br>$\{([g], \mathbb{ID}_g, m_g)\}$ | Allocation,<br>$\{([g][p][k], descr, \beta_{(g,p,k),w})\}$ | Patient type,<br>$\{([g], \mathbb{ID}_g, \hat{n}_g^{\mathrm{G}})\}$<br>Patient Sub Type,<br>$\{([g][p], \mathbb{ID}_{g,p}, \hat{n}_{g,p}^{\mathrm{GP}})\}$ |

For demonstrative purposes, a concrete example is shown below for a small fictional hospital.

**Table 6.** Concrete examples of files shown in Table 5

| scenario_1.mix | scenario_1.session | scenario_1.alloc | scenario_1.target |
|---|---|---|---|
| Case Mix,<br>[1],5<br>[2],43<br>[3],18<br>[4],9<br>[5],25<br>Sub Mix,<br>[1][1],70<br>[1][2],30<br>[2][1],100 | Patient Type,<br>[1],Specialty 1,12<br>[2],Specialty 2,25<br>[3],Specialty 3,34<br>[4],Specialty 4,10<br>[5],Specialty 5,19 | Allocation,<br>[1][1][1],Specialty 1-1@Ward 1,5.26<br>[1][2][1],Specialty 1-2@Ward 1,2.42<br>[2][1][1],Specialty 2-1@Ward 2,22.88<br>[2][1][2],Specialty 2-1@Ward 1,0<br>[2][1][3],Specialty 2-1@Ward 5,27.94<br>[3][1][1],Specialty 3-1@Ward 3,6.11<br>[3][2][1],Specialty 3-2@Ward 3,9.17<br>[3][3][1],Specialty 3-3@Ward 3,8.15<br>[4][1][1],Specialty 4-1@Ward 4,11.22 | Patient Type,<br>[1],Specialty 1,10<br>[2],Specialty 2,55<br>[3],Specialty 3,65<br>[4],Specialty 4,35<br>[5],Specialty 5,53<br>Patient Sub-Type,<br>[1][1],Specialty 1-1,5<br>[1][2],Specialty 1-2,5<br>[2][1],Specialty 2-1,55 |
| [3][1],25<br>[3][2],40<br>[3][3],35<br>[4][1],100<br>[5][1],100 | | [5][1][1],Specialty 5-1@Ward 5,0<br>[5][1][2],Specialty 5-1@Ward 4,29.38 | [3][1],Specialty 3-1,16<br>[3][2],Specialty 3-2,20<br>[3][3],Specialty 3-3,29<br>[4][1],Specialty 4-1,35<br>[5][1],Specialty 5-1,53 |

### 4.2. Task Windows and Demonstrative Examples

**4.2.1. GUI-1.** For performing a basic assessment of hospital capacity, the GUI shown in Figure 5 has been created. It has three frames. On the left are two frames of inputs that affect the analysis. The third frame on the right is the results window. The parameters that can be directly altered have a blue label to make it easier for users. The MSS template can be manipulated by altering four parameters using the scroll buttons provided. Edit boxes for these are in the top frame. After any change, the number of sessions is immediately updated (i.e., $M = weeks \times sess \times days \times |OTU|$) and displayed (i.e., edit box 5 from the top). As the number of sessions increases, the number of unassigned sessions also increases and edit box 6 from the top is automatically revised. The user should manually assign "unassigned" sessions otherwise hospital capacity is unused. The number of beds in each ward can be incremented or decremented. This requires a ward to be selected first from the drop-down combo box.

In the second frame, the number of sessions (i.e., $m_g$) assigned to a particular patient type is alterable. The default value, however, is one. Sessions can be loaded from file and alterations can be saved to file for later assessments. The "Set Even Number" button evenly assigns sessions to patient types. For instance, the total number of sessions is divided by the number of patient types. The patient case mix is not an input and is determined from the static calculations previously described in Section 3. Those calculations are performed using the ">>" button in frame three. The patient sub mix however is a requirement and should be defined in frame two manually or via the "Load Sub Mix" button. When the sub mix is directly manipulated using the scroll buttons, the proportions may no longer add to 100% The %error is shown immediately below and must be corrected before the assessment is permitted to be performed. An error message is provided to inform the user. The error can be auto corrected using the "Fix Error" button. This rescales the current percentages.

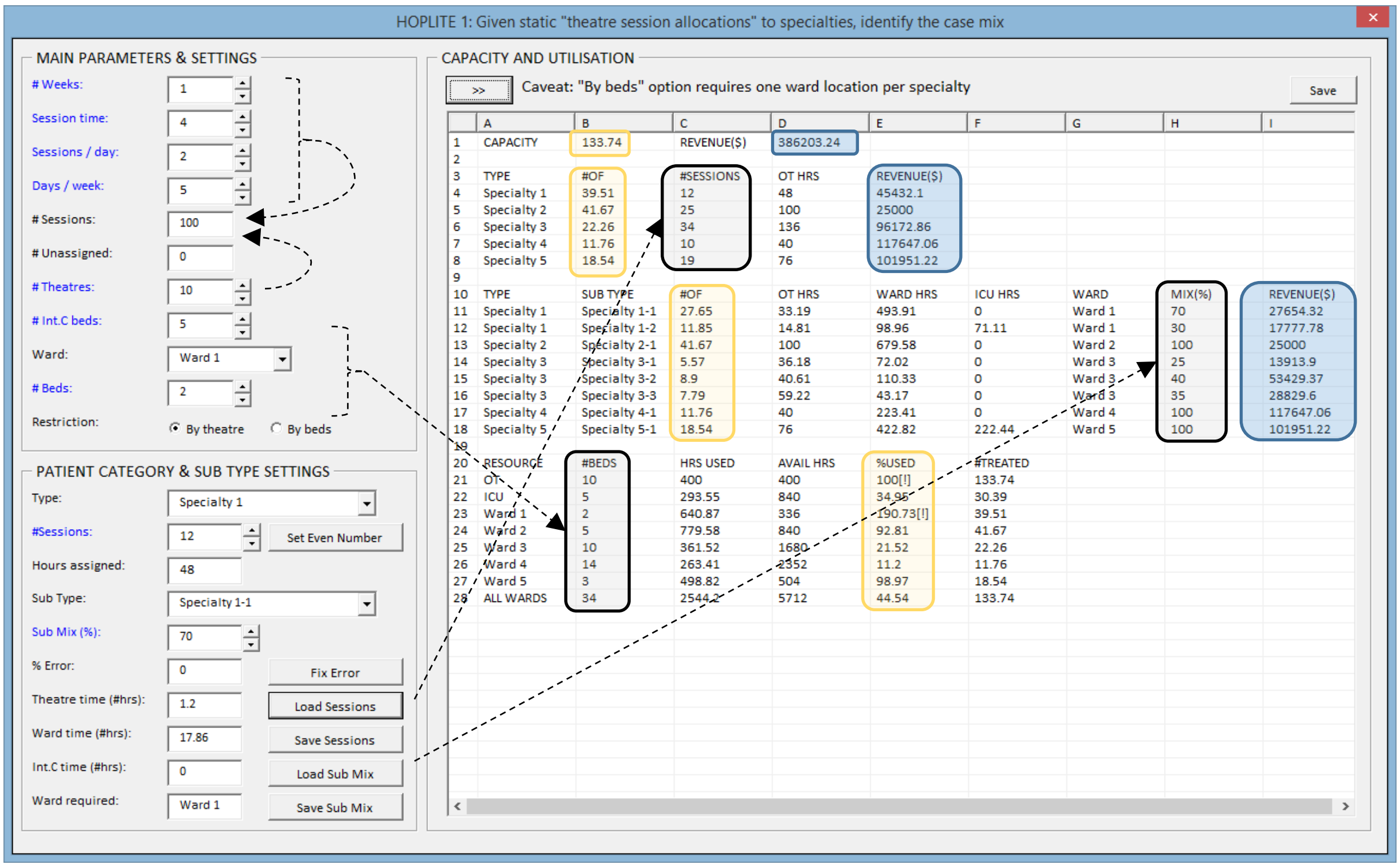

**Figure 5.** HOPLITE GUI 1 – Solution "By Theatre"

The solution report is output within a ListView with resizable columns and shows $\mathbb{N}, n^1, n^2$ and other important utilisation information; these are circled in orange. Any resource that is fully utilised or overutilised has an “[!]” next to it, so that users can see this important outcome easily. When the “By Theatre” restriction is selected, the theatre utilisation will be at 100%, whereas other resources will not. When the “By beds” restriction is used, the theatre utilisation may exceed 100% but all bed utilisations will be at 100%. These two options allow the user to consider what would happen if beds were fully saturated, or theatres and what the difference in output would look like.

**4.2.2. GUI-2.** For performing a more advanced assessment of hospital capacity, the GUI shown in Figure 6 has been created. It also has a three-frame setup like GUI-1. For this assessment a case mix and a sub mix should be selected, and the number of sessions assigned to each patient type is not required. The case mix and sub mix can be input from file or selected manually. Any % errors are automatically shown and must be corrected before analysis is permitted. The “fix” button auto-corrects the percentages by rescaling. The number of theatres and beds is an important parameter that can be altered for “what-if” analysis, that greatly affects a hospitals’ output. The MSS template parameters affect the time availability of theatres, and so must also be placed on this window.

This GUI permits users to perform a “bound” analysis for each patient type, where the maximum number of patients treatable can be determined. This is facilitated by pressing the “100%” button for a chosen patient type. Pressing this button sets the case mix for the selected patient type at 100%, while zeroing all others. The “Even” option is also provided, that partitions the case mix equally amongst the different types.

As described in Section 3, there are two case mix viewpoints. These can be selected by the radio buttons in Frame 1. There are two assessment buttons “>>” and “>>2”. These buttons activate the generation and solution of the optimization models described in Section 3.2.2. Button “>>” assumes only the first ward option is used for POSTOP. Button “>>2” permits all ward options to be considered.

The solution report shows $\mathbb{N}, n^{\mathrm{G}}, n^{\mathrm{GP}}, \beta$ and utilisation information like that shown in GUI-1; again, circled in orange. Any resource that is intended to be fully used has an “[!]” next to it. However, no resource is over-utilised as the model does not permit it.

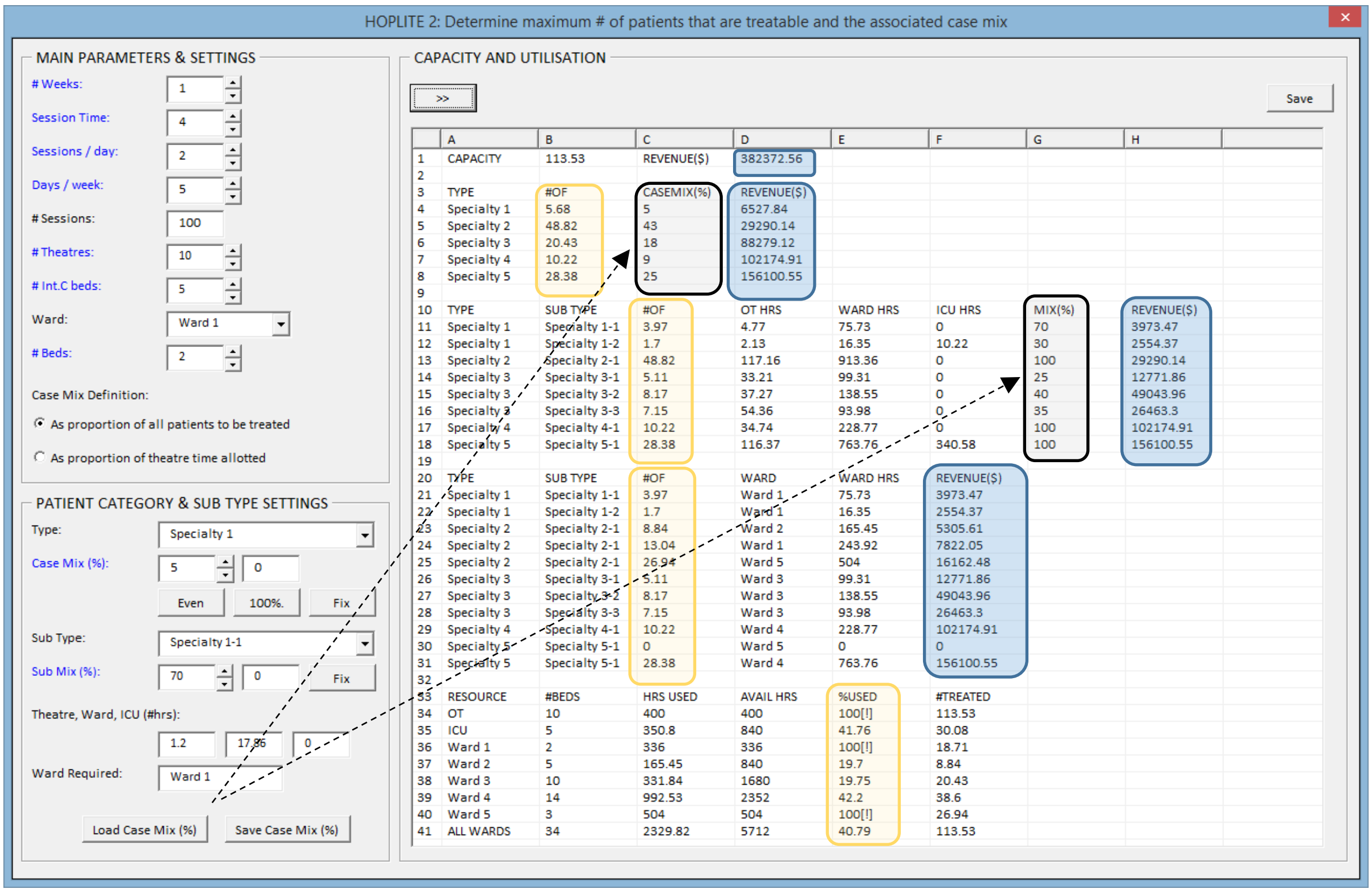

**Figure 6.** HOPLITE GUI-2. Solution given case mix definition 1

Behind the scenes, the mathematical model is generated in an excel sheet as shown in Figure 7. The setup required is evidently like the output shown in Figure 6. The capacity value shown at B1 is $\mathbb{N}$. It is the sum of the values C21:C31. These are the resource allocation (i.e., $\beta_{a,w}$) values. The value at B1 is decomposed into the values shown at B4:B8. These are the current $n_g^{\mathrm{G}}$ values. They are calculated as follows, $n_g^{\mathrm{G}} = \mu_g^{\mathrm{G}}\mathbb{N}$. They are then decomposed further using the sub mix to obtain C11:C18. The values at C11:C18 are the current values of $n_{g,p}^{\mathrm{GP}}$. These must equal the values at H11:H18 which are aggregated from C21:C31. The resource time availability constraint is imposed by setting C34:C41 to be less than or equal to D34:D41.

The values under "#ALLOCATED" and "HRS USED" can be determined using a SUMIF function call. The purpose of the SUMIF function is to aggregate specific $\beta_{a,w}$ values. An alternative approach is to write a different formula in each row, that is specific to the current $g, p$ and ward index. Both approaches have been tested, and some speed improvements have been observed if SUMIF function is not used.

| | A | B | C | D | E | F | G | H |
|---|---|---|---|---|---|---|---|---|
| 1 | CAPACITY | 113.5277 | | | | | | |
| 2 | | | | Maximize this | | | | |
| 3 | TYPE | #OF | CASEMIX(%) | | | | | |
| 4 | 1 | 5.676384 | 5 | | | | | |
| 5 | 2 | 48.8169 | 43 | | | | | |
| 6 | 3 | 20.43498 | 18 | | = constraint | | | |
| 7 | 4 | 10.21749 | 9 | | | | | |
| 8 | 5 | 28.38192 | 25 | | | | | |
| 9 | | | | | | | | |
| 10 | TYPE | PATH | #OF | OT HRS | WARD HRS | ICU HRS | MIX(%) | #ALLOCATED |
| 11 | 1 | 1 | 3.973468583 | 4.768162299 | 75.73431119 | 0 | 70 | 3.973468583 |
| 12 | 1 | 2 | 1.702915107 | 2.128643884 | 16.34798503 | 10.217491 | 30 | 1.702915107 |
| 13 | 2 | 1 | 48.81689973 | 117.1605594 | 913.3641939 | 0 | 100 | 48.81689973 |
| 14 | 3 | 1 | 5.108745321 | 33.20684458 | 99.31400903 | 0 | 25 | 5.108745321 |
| 15 | 3 | 2 | 8.173992513 | 37.27340586 | 138.5491731 | 0 | 40 | 8.173992513 |
| 16 | 3 | 3 | 7.152243449 | 54.35705021 | 93.98047892 | 0 | 35 | 7.152243449 |
| 17 | 4 | 1 | 10.21749064 | 34.73946818 | 228.7696155 | 0 | 100 | 10.21749064 |
| 18 | 5 | 1 | 28.38191845 | 116.3658656 | 763.7574254 | 340.58302 | 100 | 28.38191845 |
| 19 | | | | | | | | |
| 20 | TYPE | SUB | #OF | WARD | WARD HRS | | | |
| 21 | 1 | 1 | 3.973468583 | Ward 1 | 75.73431119 | | | |
| 22 | 1 | 2 | 1.702915107 | Ward 1 | 16.34798503 | | Decision variables | |
| 23 | 2 | 1 | 44.89577766 | Ward 2 | 840 | | | |
| 24 | 2 | 1 | 0 | Ward 1 | 0 | | | |
| 25 | 2 | 1 | 3.921122071 | Ward 5 | 73.36419394 | | | |
| 26 | 3 | 1 | 5.108745321 | Ward 3 | 99.31400903 | | | |
| 27 | 3 | 2 | 8.173992513 | Ward 3 | 138.5491731 | | | |
| 28 | 3 | 3 | 7.152243449 | Ward 3 | 93.98047892 | | | |
| 29 | 4 | 1 | 10.21749064 | Ward 4 | 228.7696155 | | | |
| 30 | 5 | 1 | 0 | Ward 5 | 0 | | | |
| 31 | 5 | 1 | 28.38191845 | Ward 4 | 763.7574254 | | | |
| 32 | | | | | | | ≤ Constraint | |
| 33 | RESOURCE | #BEDS | USED HRS | AVAIL HRS | %USED | #TREATED | | |
| 34 | OT | 10 | 400 | 400 | 100 | 113.52767 | | |
| 35 | ICU | 5 | 350.800512 | 840 | 41.76196572 | 30.084834 | | |
| 36 | Ward 1 | 2 | 92.08229621 | 336 | 27.4054453 | 5.6763837 | | |
| 37 | Ward 2 | 5 | 840 | 840 | 100 | 44.895778 | | |
| 38 | Ward 3 | 10 | 331.843661 | 1680 | 19.75259887 | 20.434981 | | |
| 39 | Ward 4 | 14 | 992.5270409 | 2352 | 42.19927895 | 38.599409 | | |
| 40 | Ward 5 | 3 | 73.36419394 | 504 | 14.55638769 | 3.9211221 | | |
| 41 | ALL WARDS | 34 | 2329.817192 | 5712 | 40.78811611 | 113.52767 | | |

**Figure 7.** Capacity model (1)

A slightly different model is required when the second case mix option is selected. For this article's small toy hospital scenario, that model is shown in Figure 8 and the results are shown in Figure 9. The output is similar but there is extra information in column C, D and E over rows 3 – 8, concerning theatre time used. Figure 9 shows that higher capacity is achieved when the case mix is used to partition MSS sessions between the different patient types. Viewing case mix relative to other patient types is more restrictive and causes a different set of bottlenecks.

| | | | | | | | | |
|---|---|---|---|---|---|---|---|---|
| 1 | CAPACITY | 134.8919 | | | Maximize this | | | |
| 2 | | | | | | | | |
| 3 | GROUP | #OF | OT HRS | AVAIL OT HRS | %USED | CASEMIX(%) | | |
| 4 | 1 | 16.46091 | 20 | 20 | 100 | 5 | | |
| 5 | 2 | 71.66667 | 172 | 172 | 100 | 43 | | |
| 6 | 3 | 11.78589 | 72 | 72 | 100 | 18 | | |
| 7 | 4 | 10.58824 | 36 | 36 | 100 | 9 | | |
| 8 | 5 | 24.39024 | 100 | 100 | 100 | 25 | | |
| 9 | | | | | | | | |
| 10 | GROUP | PATH | #OF | OT HRS | WARD HRS | ICU HRS | MIX(%) | #ALLOCATED |
| 11 | 1 | 1 | 11.52263374 | 13.82716049 | 219.6213992 | 0 | 70 | 11.52263374 |
| 12 | 1 | 2 | 4.938271605 | 6.172839506 | 47.40740741 | 29.62963 | 30 | 4.938271605 |
| 13 | 2 | 1 | 71.66666667 | 172 | 1340.883333 | 0 | 100 | 71.66666667 |
| 14 | 3 | 1 | 2.946472418 | 19.15207072 | 57.2794238 | 0 | 25 | 2.946472418 |
| 15 | 3 | 2 | 4.714355868 | 21.49746276 | 79.90833197 | 0 | 40 | 4.714355868 |
| 16 | 3 | 3 | 4.125061385 | 31.35046652 | 54.2033066 | 0 | 35 | 4.125061385 |
| 17 | 4 | 1 | 10.58823529 | 36 | 237.0705882 | 0 | 100 | 10.58823529 |
| 18 | 5 | 1 | 24.3902439 | 100 | 656.3414634 | 292.68293 | 100 | 24.3902439 |
| 19 | | | | | | | | |
| 20 | GROUP | PATH | #OF | WARD | WARD HRS | | | |
| 21 | 1 | 1 | 11.52263374 | Ward 1 | 219.6213992 | | | |
| 22 | 1 | 2 | 4.938271605 | Ward 1 | 47.40740741 | | | |
| 23 | 2 | 1 | 44.89577766 | Ward 2 | 840 | | | |
| 24 | 2 | 1 | 3.686327815 | Ward 1 | 68.97119342 | | ≤ constraint | |
| 25 | 2 | 1 | 23.08456119 | Ward 5 | 431.9121399 | | | |
| 26 | 3 | 1 | 2.946472418 | Ward 3 | 57.2794238 | | | |
| 27 | 3 | 2 | 4.714355868 | Ward 3 | 79.90833197 | | | |
| 28 | 3 | 3 | 4.125061385 | Ward 3 | 54.2033066 | | | |
| 29 | 4 | 1 | 10.58823529 | Ward 4 | 237.0705882 | | | |
| 30 | 5 | 1 | 0 | Ward 5 | 0 | | | |
| 31 | 5 | 1 | 24.3902439 | Ward 4 | 656.3414634 | | | |
| 32 | | | | | | | | |
| 33 | RESOURCE | #BEDS | USED HRS | AVAIL HRS | %USED | #TREATED | | |
| 34 | OT | 10 | 400 | 400 | 100 | 134.89194 | | |
| 35 | ICU | 5 | 322.3125565 | 840 | 38.37054244 | 29.328516 | | |
| 36 | Ward 1 | 2 | 336 | 336 | 100 | 20.147233 | | |
| 37 | Ward 2 | 5 | 840 | 840 | 100 | 44.895778 | | |
| 38 | Ward 3 | 10 | 191.3910624 | 1680 | 11.39232514 | 11.78589 | | |
| 39 | Ward 4 | 14 | 893.4120516 | 2352 | 37.98520628 | 34.978479 | | |
| 40 | Ward 5 | 3 | 431.9121399 | 504 | 85.69685316 | 23.084561 | | |
| 41 | ALL WARDS | 34 | 2692.715254 | 5712 | 47.14137349 | 134.89194 | | |

**Figure 8.** Capacity model (2)

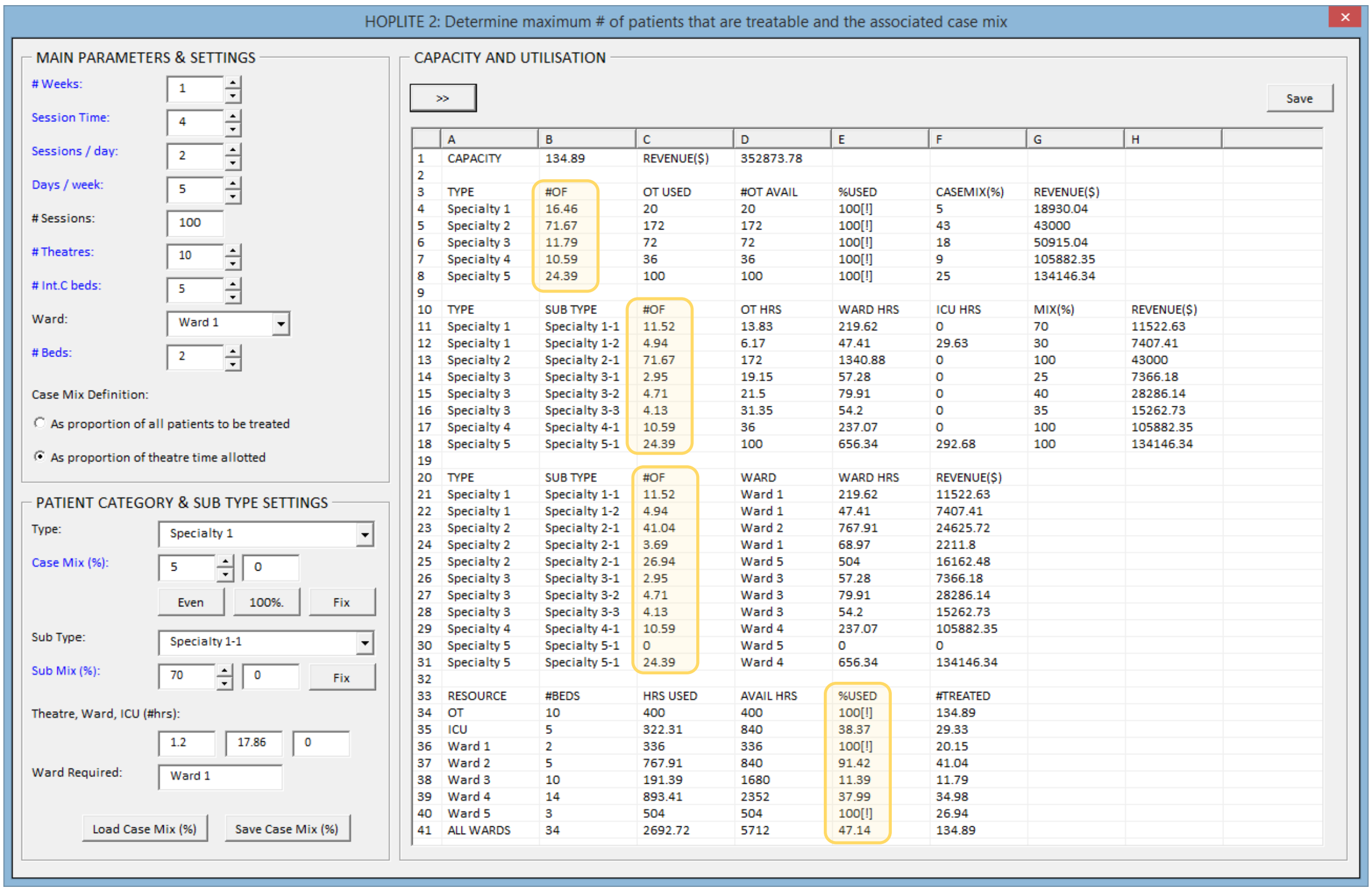

**Figure 9.** HOPLITE GUI-2. Solution given case mix definition 2

**4.2.2. GUI-3.** To evaluate a “user defined” patient cohort and resource allocation, the GUI in Figure 10 has been created.

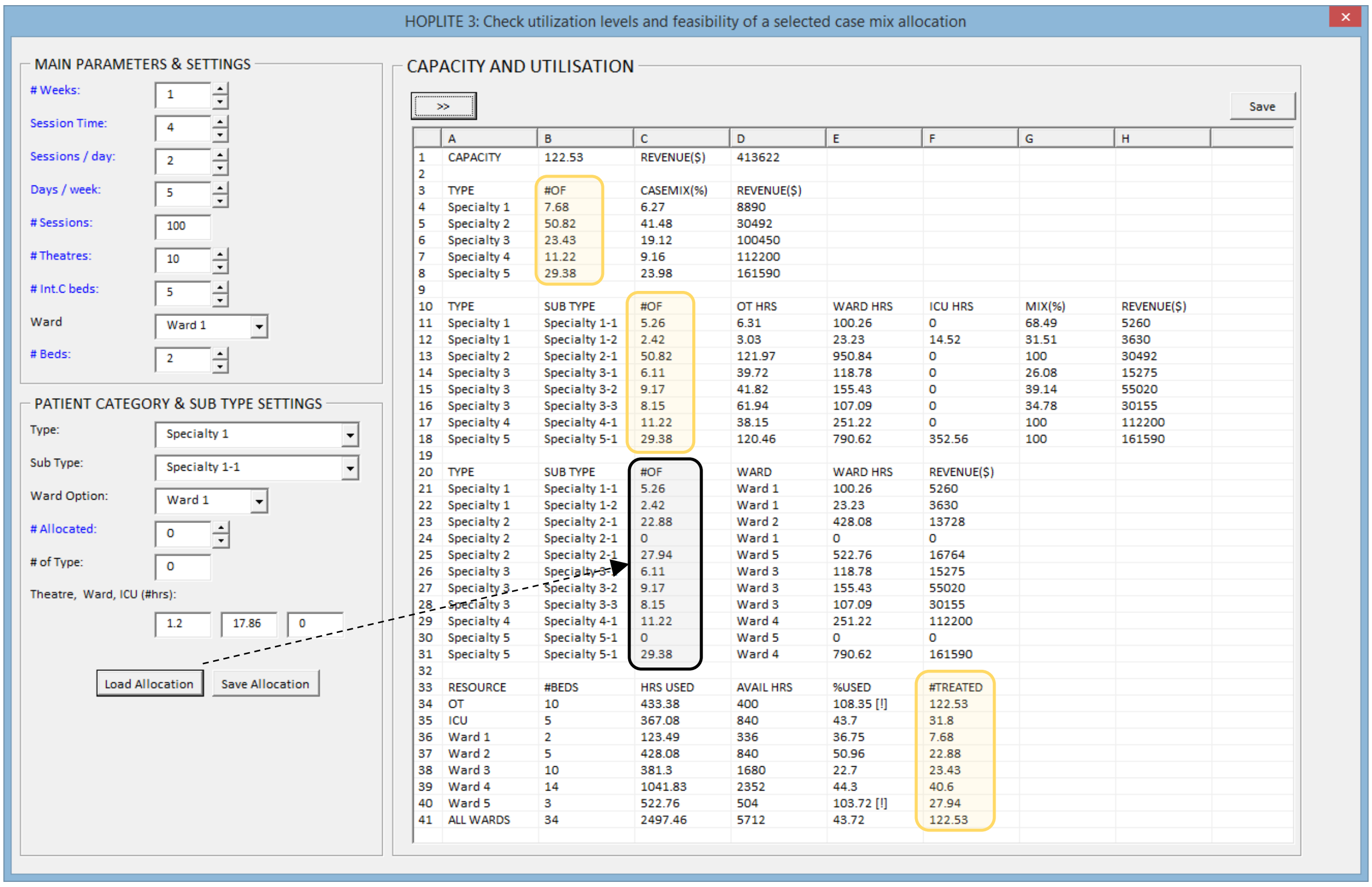

**Figure 10.** HOPLITE GUI-3. Evaluation of a case mix allocation

For the considered example, Figure 10 shows the input allocation and implied patient cohort is not quite feasible. Theatres and Ward 5 are overutilised. All other wards and the ICU are underutilised.

This means that capacity exists to treat more patients of certain types, but only if there is more theatres or theatre time.

**4.2.2. GUI-4.** To find a patient cohort that meets user defined targets, the GUI shown in Figure 11 has been created. The targets are loaded from file or else manually entered. One or both target types can be included and both difference metrics can be selected.

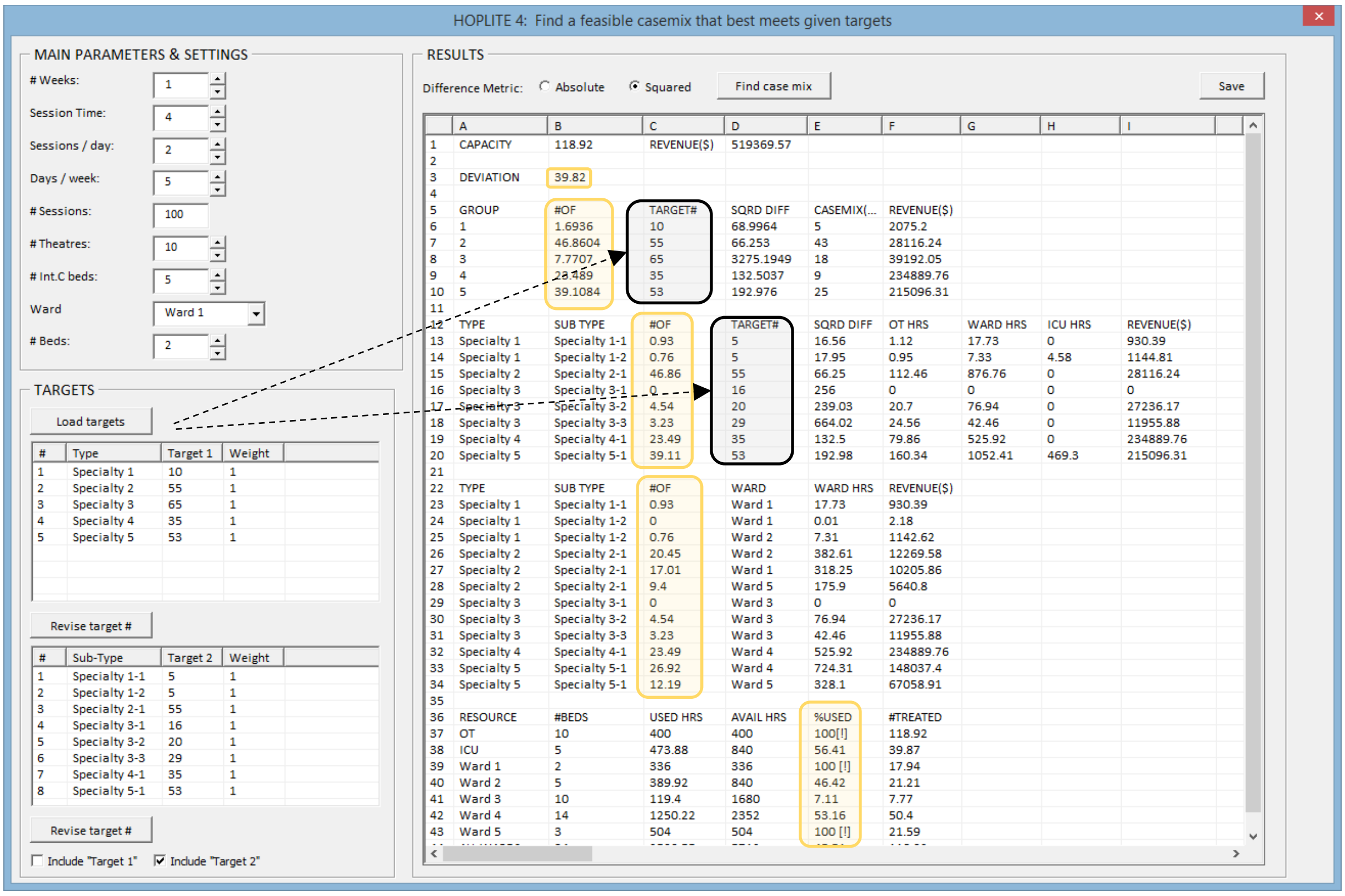

**Figure 11.** HOPLITE GUI-4. Closest case mix to selected targets

To solve the non-linear decision model, the Excel "GRG" (i.e., Generalized Reduced Gradient) non-linear solver is available. This engine was able to solve the described test problem almost immediately. The results are shown in Figure 11 for the selected $n_{g,p}^{\mathrm{GP}}$ targets. This scenario shows $n_{g,p}^{\mathrm{GP}}$ cannot be met. Other options include the non-linear solvers provided with OpenSolver. One of those is the "NOMAD" engine (i.e., see https://www.gerad.ca/nomad/) which is a "mesh adaptive direct search algorithm. This engine also solves the model but takes significantly more time than the GRG engine (i.e., approx. 3 minutes). Two other options are also available, namely COUENNE (i.e., Convex over and under envelopes for nonlinear estimation) and BONMIN (i.e., Basic Open-Source Nonlinear Mixed Integer Programming). No success however has been obtained with those to-date. The model seems to be outside the scope of those solvers.

### 4.3. Full Sized Case Study

To further demonstrate the application of HOPLITE, a full-sized case study is provided next. Our case study originates from the Princess Alexandra Hospital (PAH), Brisbane, Qld, Australia. The PAH is a large public hospital, with about one thousand beds. The PAH performs planned elective surgeries and treats acute patients surgically as they eventuate. Medical outpatients are also seen in large numbers. The hospital is spread over five floors but there are also other various "out-buildings". There is an intensive care unit (i.e., W3A & W3B with 25 beds), an emergency department, a medical assessment and planning unit, and 21 operating theatres (W3L A-E). The main wards of the PAH are summarised in Table 7. Ward W3F is a surgical care area and manages surgical patient admissions prior to surgery. W3H is a post anaesthetic care unit and provides post anaesthesia care after surgery. Ward W3D is the cardiac care ward, also known as WCARD, and W3E is the cardiac care unit ward, commonly called WCCU. Radiology and imaging are performed in W1F, W1H, W1L.

**Table 7.** Main recovery wards

| WARD | # BEDS | WARD | # BEDS |
|---|---|---|---|
| W3B | 10 | W1C, W2B, W5B, W5C, W5D | 24 |
| W4BR, WCARD | 14 | W1D, W4E | 26 |
| W3A | 15 | W2A, W2D, W3C, W4C, W4D, W5A | 28 |
| W4BT | 16 | W2E | 29 |
| W4A | 19 | W2C | 40 |
| W1A, W1B, W3D, W3E, W3F, W3H, WTRANS, W3D, W3E | 20 | | |

De-identified patient data has been collected from the PAH. From that data, patient types are defined relative to the specialties shown in Table 8. Patient sub types are categorised by the Australian Refined Diagnosis Related Groups (AR-DRGs) relevant to each of the twenty-one specialties. The Australian Refined Diagnosis Related Groups (AR-DRGs) (https://www.ihpa.gov.au/what-we-do/ar-drg-classification) is a classification system, which provides a clinically meaningful way to relate the number and type of patients treated in a hospital to the resources required by the hospital. AR-DRGs group patients with similar diagnoses requiring similar hospital services.

The exact number of sub types is also shown in Table 8. In total there are 338. We have reduced the overall number of sub types, by aggregating the "A", "B", "C", and "D" AR-DRG variants that exists. The following is an example of three variants that are considered as one sub type:

B66A M Nervous System Neoplasms W Radiotherapy
B66B M Nervous System Neoplasms W/O Radiotherapy W Catastrophic or Severe CC
B66C M Nervous System Neoplasms W/O Radiotherapy W/O Catastrophic or Severe CC

Otherwise, there would be a total of 807 DRG's which is prohibitive. The classification of patient sub type as surgical, medical, or other is shown in brackets, i.e. (15|8|0) indicates 15 surgical and 8 medical types.

**Table 8.** Patient types

| | TYPE | #SUB | | TYPE | # SUB | | TYPE | # SUB |
|---|---|---|---|---|---|---|---|---|
| **1** | Cardiology | 23 (15\|8\|0) | 8 | Hepatology | 13 (6\|5\|2) | 15 | Otolaryngology | 16 (9\|7\|0) |
| **2** | Endocrinology | 17 (10\|5\|2) | 9 | Immunology | 8 (1\|6\|1) | 16 | Plastics | 18 (10\|8\|0) |
| **3** | Dental | 2 (0\|1\|1) | 10 | Neurology | 33 (7\|23\|3) | 17 | Psychiatry | 10 (0\|9\|1) |
| **4** | Faciomaxillary | 3 (2\|0\|1) | 11 | Nephrology | 20 (8\|9\|3) | 18 | Respiratory | 22 (2\|17\|3) |
| **5** | Gastroenterology | 20 (10\|7\|3) | 12 | Oncology | 8 (4\|4\|0) | 19 | Transplants | 11 (10\|0\|1) |
| **6** | Gynaecology | 13 (10\|3\|0) | 13 | Ophthalmology | 16 (12\|4\|0) | 20 | Urology | 12 (6\|5\|1) |
| **7** | Hematology | 5 (2\|3\|0) | 14 | Orthopaedic | 51 (29\|21\|1) | 21 | Vascular | 17 (6\|7\|4) |

The resource consumption profiles and revenues associated with each patient sub type have been extracted and can be found in our supplementary data document. All revenues are defined relative to the "National Efficient Price" (NEP). The NEP is the Australian national price for the average cost of public hospital activity (i.e., 1 NWAU). The NEP was $5134 during 2019-2020. (Health Funding Policy and Principles 2019-20).

***4.3.1. Part 1***. In our case study an advanced assessment of capacity was first performed. The specified case mix is shown in Table 9 and the sub mix is shown in Table 10. For the capacity assessment, a period of 4 weeks was selected. As such the master surgical schedule has 840 sessions (i.e., two/day × five days/week × four weeks × 21 theatres).

**Table 9.** Patient case mix

| | TYPE | % | | TYPE | % | | TYPE | % |
|---|---|---|---|---|---|---|---|---|
| **1** | Cardiology | 6.765 | 8 | Hepatology | 3.824 | 15 | Otolaryngology | 5.294 |
| **2** | Endocrinology | 5 | 9 | Immunology | 2.353 | 16 | Plastics | 5.294 |
| **3** | Dental | 0.588 | 10 | Neurology | 9.706 | 17 | Psychiatry | 2.941 |
| **4** | Faciomaxillary | 0.882 | 11 | Nephrology | 5.882 | 18 | Respiratory | 6.471 |
| **5** | Gastroenterology | 5.882 | 12 | Oncology | 2.353 | 19 | Transplants | 3.235 |
| **6** | Gynaecology | 3.824 | 13 | Ophthalmology | 4.706 | 20 | Urology | 3.529 |
| **7** | Hematology | 1.471 | 14 | Orthopaedic | 15 | 21 | Vascular | 5 |

**Table 10.** Patient sub mix

| TYPE (#SUB) | SUB MIX |
|---|---|
| Cardiology (23) | 2.25,0.76,6.02,4.53,8.72,8.48,6.06,4.37,7.03,1.05,0,0,2.02,0.1,7.41,1.97,6.51,2.26,6.76,3.74,10.12,8.34,1.5 |
| Endocrinology (17) | 7.28,7.35,3.45,9.06,2.61,2.99,3,0,9.55,5.34,7.51,10.26,9.24,9.82,5.1,2.37,5.07 |
| Dental (2) | 55.75,44.25 |
| Faciomaxillary (3) | 70.67,0,29.33 |
| Gastroenterology (20) | 8.28,5.8,2.26,6.85,3.5,0,7.43,6.89,5.68,8.28,1.93,1.03,3.54,7.13,7.67,1.75,6.24,4.55,2.94,8.25 |
| Gynaecology (13) | 0,4.53,0,11.36,3.36,4.95,10.78,15.17,17.3,0,17.6,3.55,11.4 |
| Hematology (5) | 37.33,2.56,26.4,23.29,10.42 |
| Hepatology (13) | 5.37,4.62,13.43,14.32,6.68,1.55,0,0,0.51,0.01,19.88,11.11,22.52 |
| Immunology (8) | 0,5.66,20.95,4.78,18.86,10.16,15.65,23.94 |
| Neurology (33) | 6.11,4.05,5.84,3.72,5.38,1.39,0.46,0,3.85,3.59,0,0,0,2.99,1.2,0,4.18,6.24,3.72,5.64,4.18,6.57,0.6,6.04,0,0,6.11, 3.52,1.33, 0.4,6.31,2.86,3.72 |
| Nephrology (20) | 9.61,5.61,1.03,4.83,1.39,1.49,4.11,0.23,9.21,8.33,8.34,0.27,10.73,1.81,4.74,3.12,1.72,9.26,11.81,2.36 |
| Oncology (8) | 16.49,20.59,16.28,3.92,2.19,23.11,13.66,3.76 |
| Ophthalmology (16) | 6.77,8.44,6.31,4.84,9.39,0.98,4.55,6.97,6.22,5.13,5.51,3.72,8.76,7.77,4.42,10.22 |
| Orthopaedic (51) | 3.2,0,0.49,0.4,1.45,2.63,2.94,1.9,3.3,0.14,0.47,3.17,3.09,1.35,2.6,2.78,1.55,3.36,2.57,2,3.03,2.92,2.39, 2.5,1.44,3.33,2.42,3.5,3.08,0,0.28,1.37,2.46,2.86,1.17,2.56,3.52,1.61,1.87,2.6,1.39,1.61,0.09,0.43,2.25,1.78,2.04,1.78,2.31,2.02,0 |
| Otolaryngology (16) | 0,10.42,1.72,12.09,7.44,8.77,3.5,2.99,7.15,2.87,0.39,8.78,8.01,15.62,7.43,2.82 |
| Plastics (18) | 4.9,2.71,9.4,7.98,3.09,6.97,8.3,9.4,9.9,3.04,3.13,8.23,0.79,9.02,1.33,8.88,0.11,2.82 |
| Psychiatry (10) | 4.55,3.58,8.7,15.42,15.21,14.58,5.24,0.93,20.18,11.61 |
| Respiratory (22) | 4.84,0.78,4.21,6.7,8.01,0,0.39,8.16,1.92,3.56,8.38,2.96,7.15,1.97,6.2,6.06,8.26,0,6.13,4,5.37,4.95 |
| Transplants (11) | 0,0,0,0,0,0,0,0,0,0,100 |
| Urology (12) | 4.41,14.07,8.04,2.33,2.22,12.66,15.34,6.63,12.39,10.93,0,10.98 |
| Vascular (17) | 8.52,5.27,3.72,1.67,2.49,10.18,4.11,10.44,11.58,2.6,11.67,6.57,4.17,1.15,3.48,8.45,3.93 |

To perform the assessment, the two methods from Section 3.3.2 were applied. The time to solve the model and report the results was quite fast and was completed in under a minute. This task, however, was not completed instantaneously. Before OpenSolver can be activated, Excel requires time to extract and copy the model to file.

The main results are summarised in Figure 12 and Figure 13 for the two case mix viewpoints, and a comparison of the differences is shown in Figure 14. In Figure 15 and Figure 16, the ward utilisation statistics are shown. For the first option, 921 patients are achievable, and this amounts to about 11050 over the course of a year. For the second option, the output is significantly higher with 1864 patients per month, and 22368 over the course of a year. The reason for the difference, is due to how trade-offs between patient types is managed. The first option is more restrictive as we have already discussed. Under the second case mix definition, the model is permitted to choose whether certain patient types are treated or not. In this case study, an increased number of patients can be achieved overall by "zeroing" some of the patient types. For instance, in our solution, we can see that no cardiology patients were selected, even though there was free theatre time allocated to treat them. Under the first case mix definition, this is not possible, and cardiology patients are selected. Choosing no cardiology patients is not realistic and highlights a minor defect/quirk of the second case mix definition. That quirk however can be overcome by adding a minimum requirement for each patient type (a.k.a., a minimum demand or target). Consequently, some additional constraints should be added to the model when using the second case mix option.

The revenues associated with the obtained patient case mixes are in the order of 17 to 24 million dollars per month. Based upon this data, the total revenue over the course of one year would lie in the region of AUD 200-300 million.

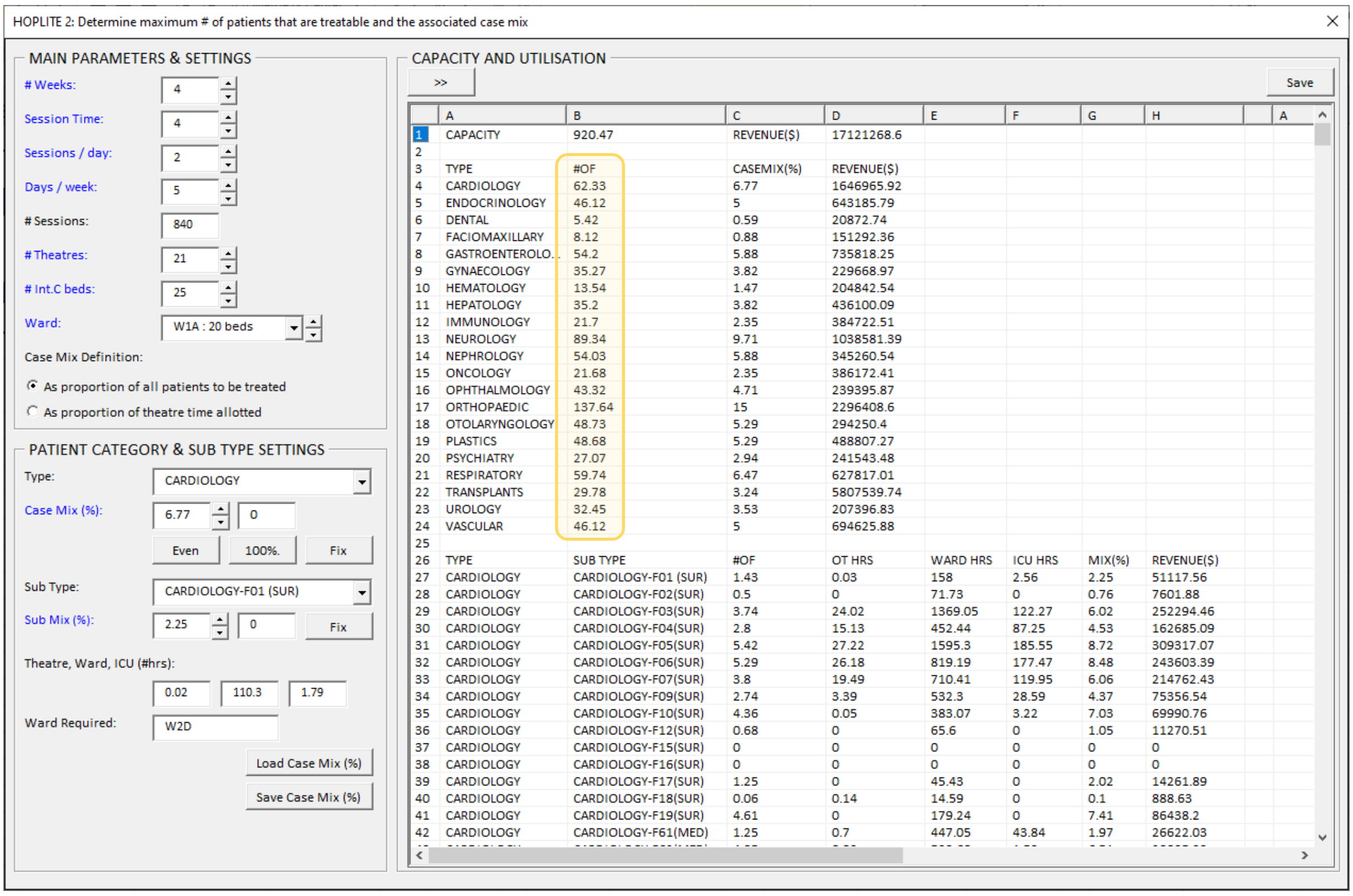

**Figure 12**. Capacity assessment results (Case mix option 1)

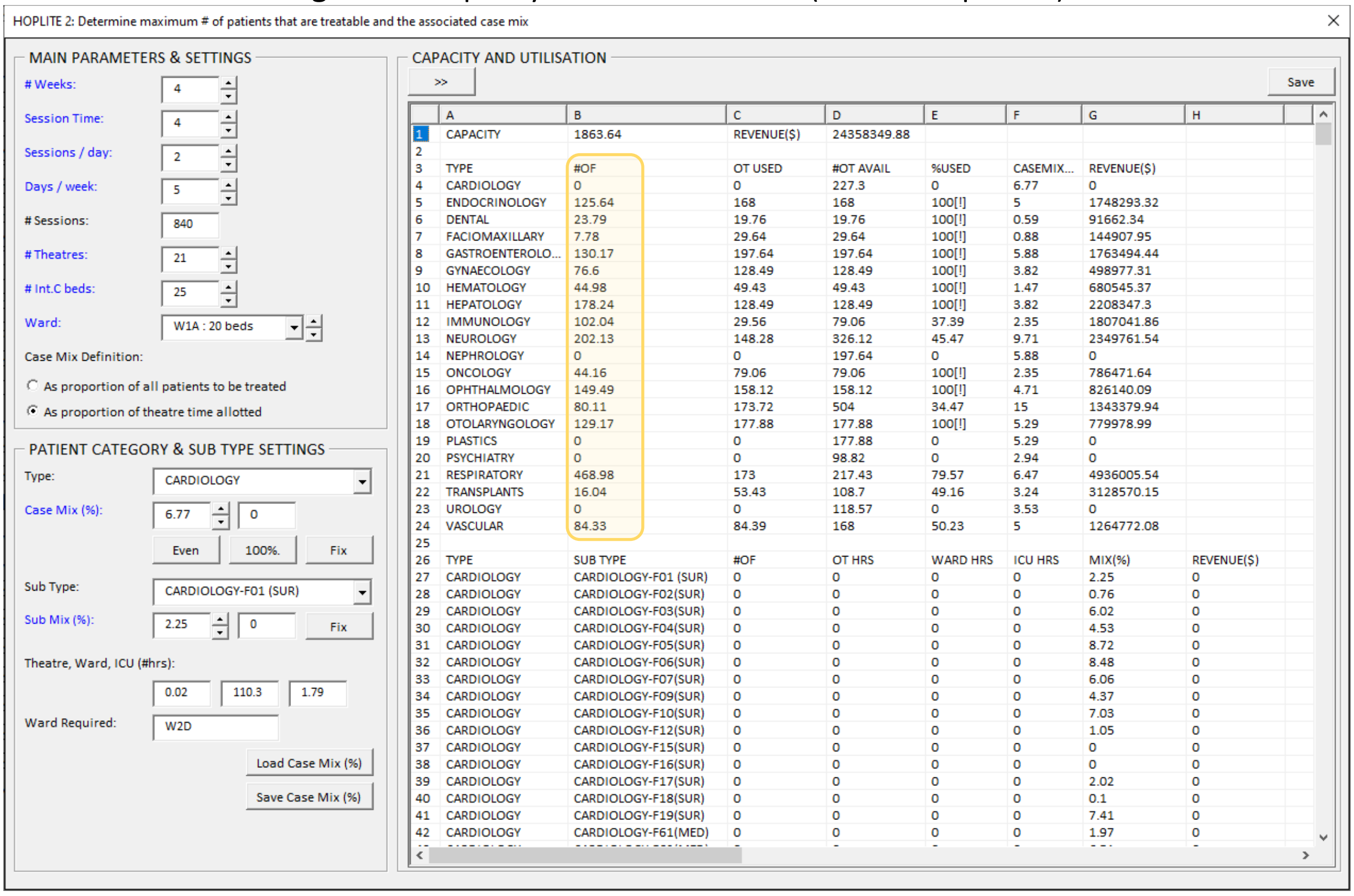

**Figure 13.** Capacity assessment results (Case mix option 2)

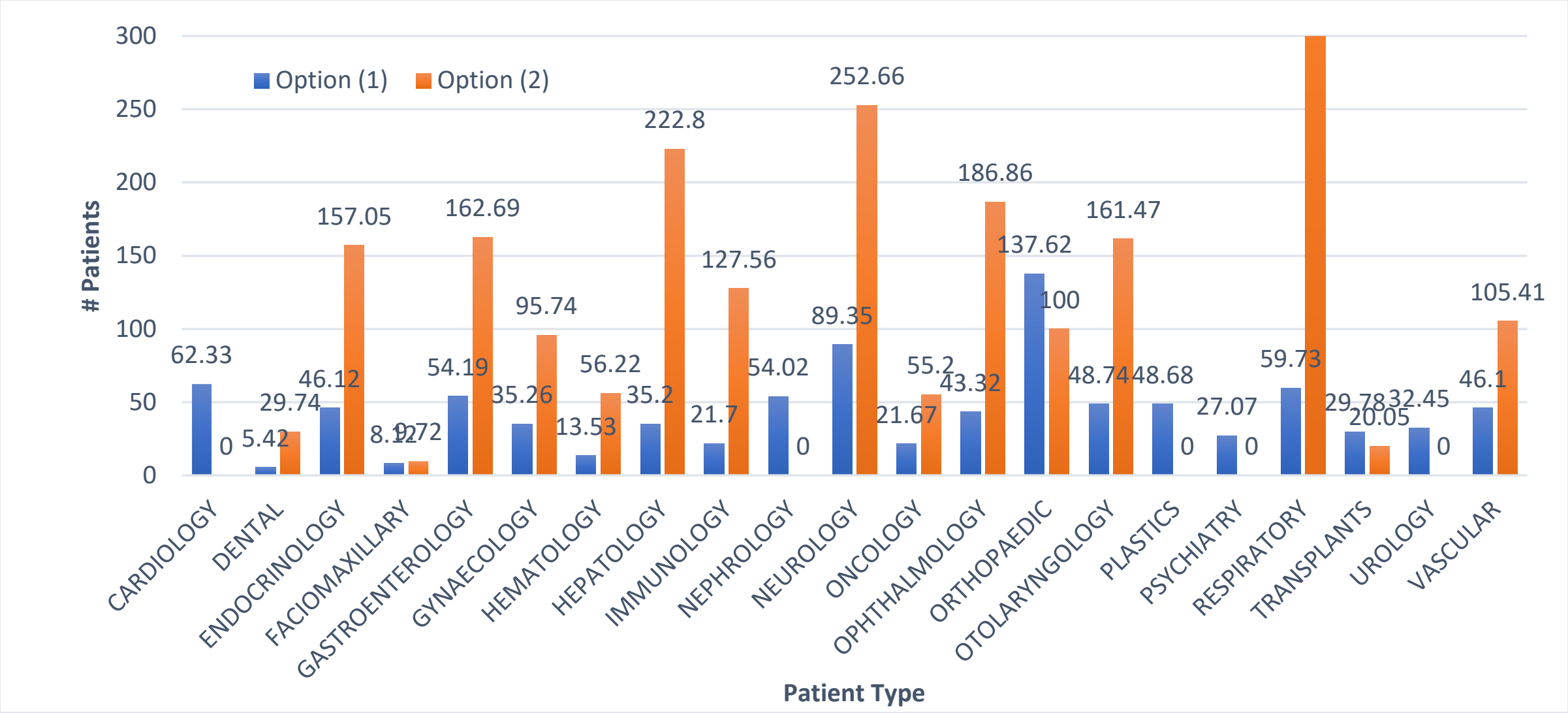

**Figure 14.** Comparison of patient cohorts identified

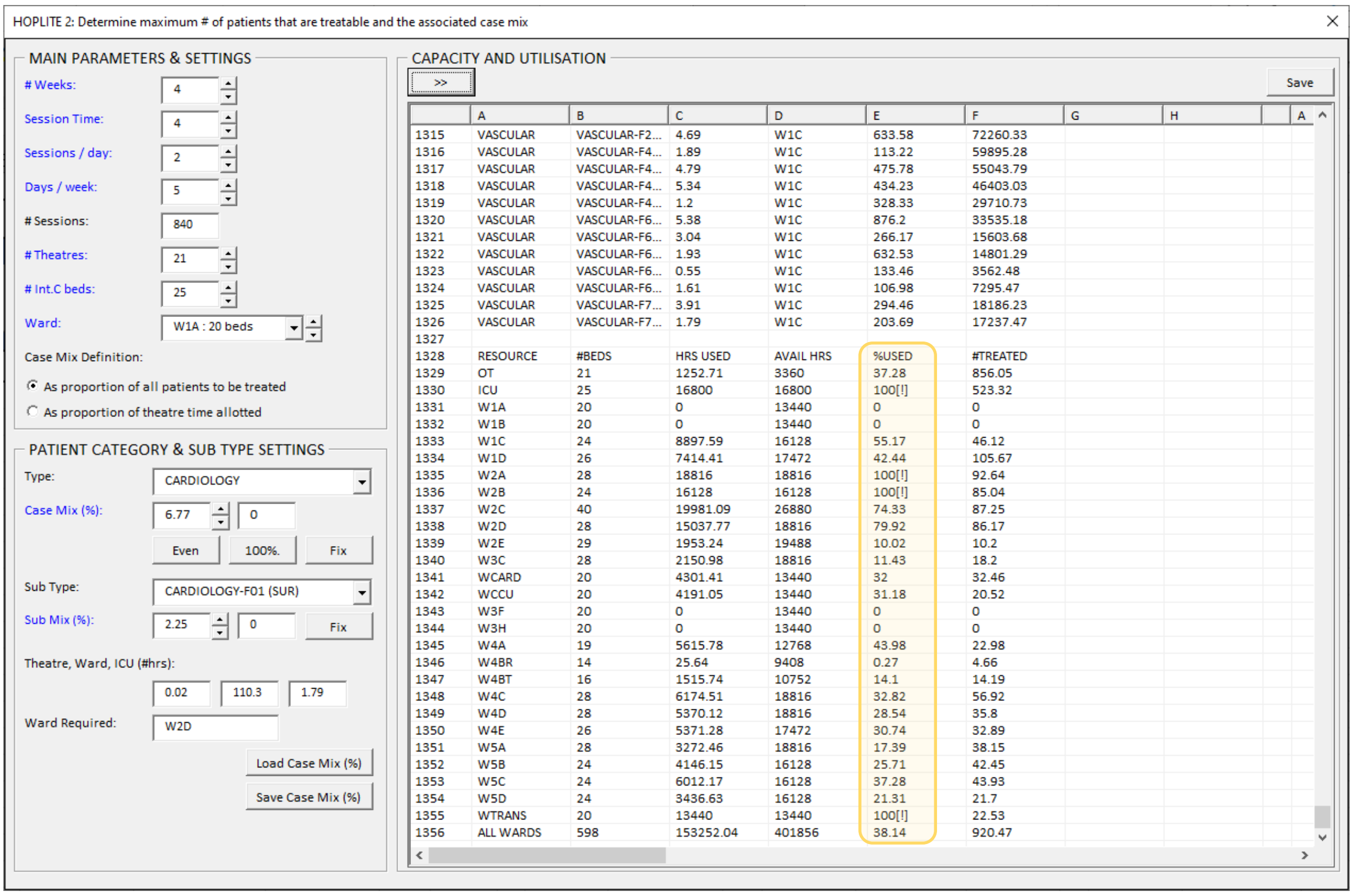

**Figure 15.** Ward utilisations (Case mix option 1)

In Figure 15, we can see that only four of the wards are fully occupied. These wards are bottlenecks and restrict further outputs. The ICU is one of them. Most of the other wards are also moderately utilised. In Figure 16, we can see that most of the wards are fully utilised. In both scenarios, the theatres are not deemed to be bottlenecks. Also, a few wards have zero utilisation. This occurs because the current patient types do not have those wards as candidate locations for medical or surgical care.

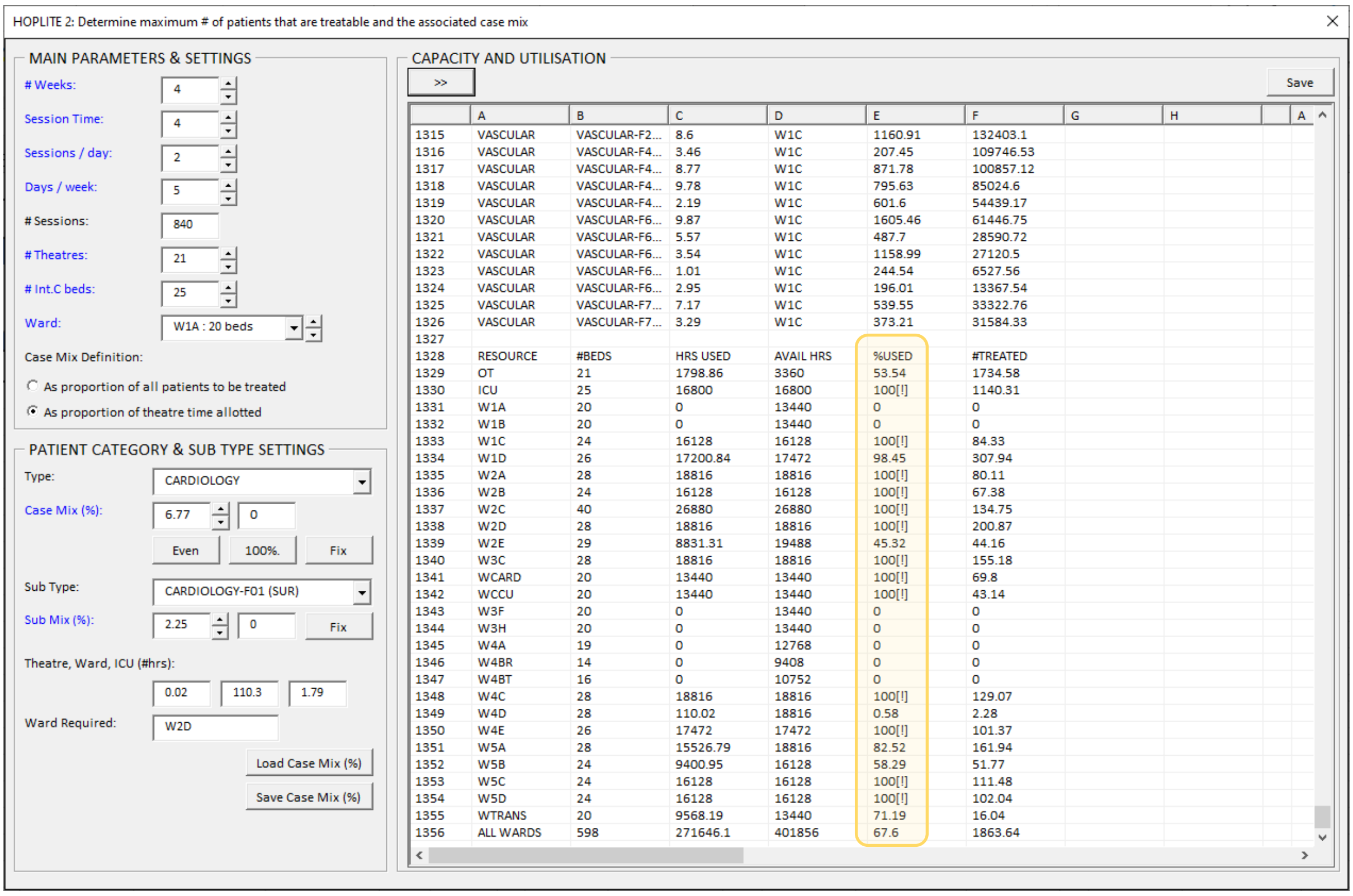

**Figure 16**. Ward utilisations (Case mix option 2)

***4.3.2 Part 2.*** A best-fit case mix identification task was performed next. For this analysis, a target cohort of an appropriate nature was defined arbitrarily. Specific targets $\hat{n}_g^G$ are defined for each patient type, and these are displayed in Figure 17.

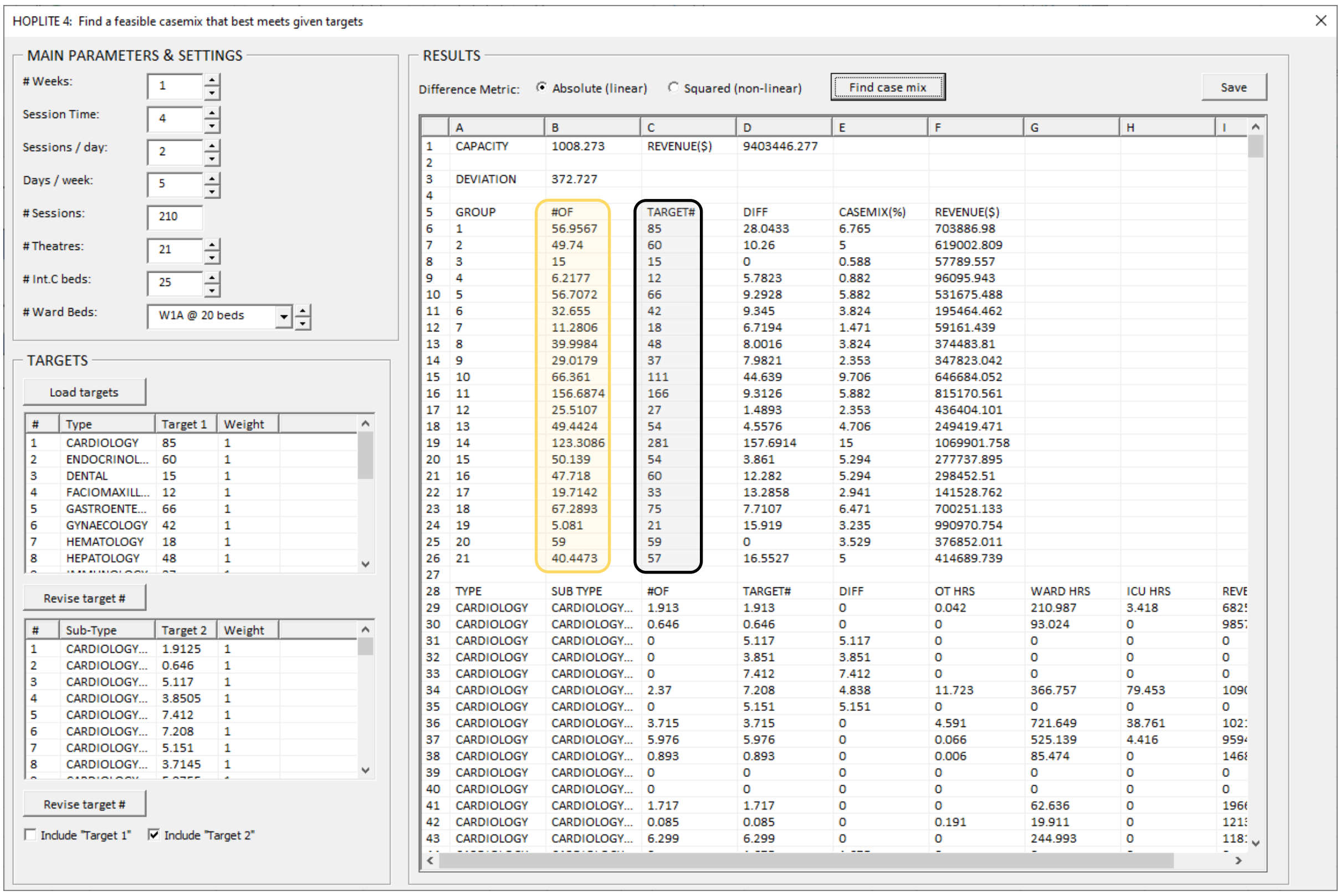

**Figure 17.** Best fit case mix (one month)

Patient sub type targets were also defined, according to $\hat{n}_{g,p}^{\mathrm{GP}} = \mu_{g,p}^{\mathrm{GP}} \hat{n}_{g}^{\mathrm{G}}$. These are listed explicitly in our supplementary data document. The feasibility of that patient cohort for a one-month period was queried by solving the model from section 3.2.4. The best fit case mix is shown in Figure 17. The specified case mix exceeds the capacity of the hospital, and all the wards, theatres, and intensive care unit are at 100% utilisation. The targets constitute a total of 1381 patients but the maximum number of patients that are treatable is 1008.27. The targets cannot be met without reconfiguration of the hospital, or an increase in the period considered. Table 11 shows how the targets are gradually met as the number of weeks is increased.

**Table 11.** Best fit case mixes

| GROUP | TARGET# | #TREATED | | | | #UNMET | | | |
|---|---|---|---|---|---|---|---|---|---|
| | | 1wk | 2wks | 3wks | 4wks | 1wk | 2wks | 3wks | 4wks |
| **1** | 85 | 56.9567 | 85 | 85 | 85 | 28.0433 | 0 | 0 | 0 |
| **2** | 60 | 49.74 | 60 | 60 | 60 | 10.26 | 0 | 0 | 0 |
| **3** | 15 | 15 | 15 | 15 | 15 | 0 | 0 | 0 | 0 |
| **4** | 12 | 6.2177 | 12 | 12 | 12 | 5.7823 | 0 | 0 | 0 |
| **5** | 66 | 56.7072 | 66 | 66 | 66 | 9.2928 | 0 | 0 | 0 |
| **6** | 42 | 32.655 | 42 | 42 | 42 | 9.345 | 0 | 0 | 0 |
| **7** | 18 | 11.2806 | 18 | 18 | 18 | 6.7194 | 0 | 0 | 0 |
| **8** | 48 | 39.9984 | 48 | 48 | 48 | 8.0016 | 0 | 0 | 0 |
| **9** | 37 | 29.0179 | 37 | 37 | 37 | 7.9821 | 0 | 0 | 0 |
| **10** | 111 | 66.361 | 91.4824 | 105.4887 | 111 | 44.639 | 19.5176 | 5.5113 | 0 |
| **11** | 166 | 156.6874 | 166 | 166 | 166 | 9.3126 | 0 | 0 | 0 |
| **12** | 27 | 25.5107 | 27 | 27 | 27 | 1.4893 | 0 | 0 | 0 |
| **13** | 54 | 49.4424 | 54 | 54 | 54 | 4.5576 | 0 | 0 | 0 |
| **14** | 281 | 123.3086 | 177.0855 | 208.492 | 231.1368 | 157.6914 | 103.9145 | 72.508 | 49.8632 |
| **15** | 54 | 50.139 | 54 | 54 | 54 | 3.861 | 0 | 0 | 0 |
| **16** | 60 | 47.718 | 60 | 60 | 60 | 12.282 | 0 | 0 | 0 |
| **17** | 33 | 19.7142 | 33 | 33 | 33 | 13.2858 | 0 | 0 | 0 |
| **18** | 75 | 67.2893 | 75 | 75 | 75 | 7.7107 | 0 | 0 | 0 |
| **19** | 21 | 5.081 | 8.5458 | 17.2985 | 21 | 15.919 | 12.4542 | 3.7015 | 0 |
| **20** | 59 | 59 | 59 | 59 | 59 | 0 | 0 | 0 | 0 |
| **21** | 57 | 40.4473 | 54.1819 | 57 | 57 | 16.5527 | 2.8181 | 0 | 0 |
| | Sum | 1008.2724 | 1242.2956 | 1299.279 | 1331.1368 | 372.7276 | 138.7044 | 81.7208 | 49.8632 |

**Final Remarks.** During development, we have observed that opening the Solver or OpenSolver window from the ribbon menu in Excel is not always instantaneous, particularly when instances with many patient types and sub types are loaded. Smaller instances and those of the scale of our case study showed no such issue.

Finding a best-fit case mix is the hardest task currently facilitated by HOPLITE. The free NLP solvers do not seem to be capable of solving large instances like our case study with a sum of squares objective. The linear objectives described by equation (21) and (23), however, provide no difficulty at all. The non-linear sum of squares objective, however, can be managed better in one of two ways. We have previously shown that the sum of squares objective is a quadratic function. As such, a Quadratic Programming solver present for instance in a commercial package like CPLEX, can be applied. Unfortunately, OpenSolver does not have such an approach presently. Another approach worth considering is Separable Programming. The sum of squares objective can be approximated with arbitrary precision as a piecewise linear function. OpenSolver is certainly capable of solving the resulting mixed integer programming formulation.

When targets $\hat{n}_{g}^{\mathrm{G}}$ are chosen, the optimization model of Section 3.2.4 will choose the sub mix $n_{g,p}^{\mathrm{GP}}$ in an unrestricted way. In other words, some $n_{g,p}^{\mathrm{GP}}$ will be zeroed, in favour of other sub types $n_{g,p'}^{\mathrm{GP}}$, within a patient type grouping. If this is unwanted, then both target types should be included in the objective.

## 5. Conclusions and Managerial Insights

This article demonstrates how an Excel-based personal decision support tool (PDST) called HOPLITE was designed and how it can help hospital managers perform capacity assessments and other capacity related queries. This article demonstrates how theory can be put into practice and describes the practicalities of creating a graphical user interface, inputting and outputting information, performing optimizations, applying quantitative methods, and reporting results.

HOPLITE has many capabilities and is necessary for a variety of reasons. The PDST can reduce the workload of planning staff. It may be used to inform hospital managers and other planners of the decisions they can make to avert future problems (Kreuger 2018). It can be used to make judgements about a hospitals' capability, in the future, to treat cohorts of different patient types. Our PDST specifically identifies patient case mixes that meet certain guidelines and constraints, or evaluates those that have been defined elsewhere, by some other technique. The development process has shown a necessity to link the master surgical schedule template to the PDST. Otherwise, assessments and evaluations are restricted and lack realism.

After significant development and testing, the main limitation of our approach is the capability for Excel to quickly generate a model with tens of thousands of variables. For the most part, the free solvers are adequate, and quick once a model has been generated. HOPLITE is best suited for small and medium sized instances at this point, and further development is needed to handle instances arising from larger hospitals, or where more fine-grained analysis, resulting in more patient types and sub types, is needed. These weaknesses, however, are not expected to exist if a commercial optimization software is integrated, but this requires a subscription or a licencing agreement.

HOPLITE is a minimal viable product, and it is intended that the software will continue to be developed to provide further functionality. Because it is written in VBA, the tool can be easily extended with additional functionality. In future it is worth considering whether a translator that converts raw hospital data into a format usable for hospital capacity assessment should be created. The existence of a translator motivates the creation of a central database to hold all relevant information. Currently HOPLITE does not use any type of database and information is placed in separate text files. The joint usage of a translator and database may facilitate an easier integration of the software to a hospital's IT system. The database design that should be implemented, however, is specific to the needs of different hospitals and is a topic for a future study.

Some simplifying assumptions have been made, to facilitate the development of a viable prototype PDST. Our tool only includes the primary resources of a hospital, like theatres, wards, and intensive care beds. The quantitative techniques, however, are designed with the capability to handle additional resources. As such, a more general PDST could be developed. However, we believe that it is necessary to await further positive feedback by hospitals, and their staff, before new extensions are considered, and current assumptions are removed.

Regarding the uptake of a tool like HOPLITE, the availability of data weighs heavily. Whether hospitals can provide free flow of the necessary information and knowledge is debatable, however we believe the answer is ultimately yes, and probably in a not-too-distant future. This PDST can be largely used offline since historical data can be prepared and uploaded into a PDST project file. Long term patterns in case-mix and resource utilisation change infrequently, and so only occasional updates to historical case mix data would be required. This means that costly direct integration with existing IT systems is not necessary for this tool to be used. The deployment of a PDST like HOPLITE is expected to be beneficial, however there is no guarantee. Whether the PDST can improve decision making outcomes and decision-making processes, is an open question, we hope can be answered.

**Funding**. This research was funded by the Australian Research Council (ARC) Linkage Grant LP 180100542 and supported by the Princess Alexandra Hospital and the Queensland Children's Hospital in Brisbane, Australia. Thanks to Dr Andy Wong for assisting with data extraction activities related to this articles' case study.

**Appendix A-1.** Let $Z = \sum_g \left(n_g^{\mathrm{G}} - \hat{n}_g^{\mathrm{G}}\right)^2$ . By expansion, $Z = \sum_g \left(n_g^{\mathrm{G}}\right)^2 - 2\sum_g \hat{n}_g^{\mathrm{G}} n_g^{\mathrm{G}} + \sum_g \left(\hat{n}_g^{\mathrm{G}}\right)^2$.
**Property**: $Z$ is a quadratic function, and as such, $Z = \frac{1}{2}x^{\mathrm{T}}Hx - \Phi^{\mathrm{T}}x + c$ where:

- $x = \begin{bmatrix} n_1^{\mathrm{G}} & n_2^{\mathrm{G}} & \dots & n_{|G|}^{\mathrm{G}} \end{bmatrix}^{\mathrm{T}}$ (i.e., the vector of decisions)
- $\Phi = \begin{bmatrix} 2\hat{n}_1^{\mathrm{G}} & 2\hat{n}_2^{\mathrm{G}} & \dots & 2\hat{n}_{|G|}^{\mathrm{G}} \end{bmatrix}^{\mathrm{T}}$ (i.e., the vector of targets multiplied by two)

- $c = \sum_g \left(\hat{n}_g^{\mathrm{G}}\right)^2$ (i.e., the sum of the targets squared)
- $H = 2I_{|G|}$ is the Hessian of the function $f = \sum_g \left(n_g^{\mathrm{G}}\right)^2, f: \mathbb{R}^{|G|} \to \mathbb{R}$.
- Diagonals of $H$ are $\frac{\partial^2 f}{\partial n_g^{\mathrm{G}}} = 2$ and non-diagonals are $\frac{\partial^2 f}{\partial n_g^{\mathrm{G}} \partial n_{g'}^{\mathrm{G}}} = 0 \ \forall g, g' | g \neq g'$

**Example:** Given, $|G| = 3$ and $\hat{n}_1^{\mathrm{G}} = 10, \hat{n}_2^{\mathrm{G}} = 40, \hat{n}_3^{\mathrm{G}} = 20$ then:

$$Z = \frac{1}{2} x^{\mathrm{T}} H x - \Phi^{\mathrm{T}} x + c = \frac{1}{2} \begin{bmatrix} n_1^{\mathrm{G}} & n_2^{\mathrm{G}} & n_3^{\mathrm{G}} \end{bmatrix} \begin{bmatrix} 2 & 0 & 0 \\ 0 & 2 & 0 \\ 0 & 0 & 2 \end{bmatrix} \begin{bmatrix} n_1^{\mathrm{G}} \\ n_2^{\mathrm{G}} \\ n_3^{\mathrm{G}} \end{bmatrix} - \begin{bmatrix} 20 & 80 & 40 \end{bmatrix} \begin{bmatrix} n_1^{\mathrm{G}} \\ n_2^{\mathrm{G}} \\ n_3^{\mathrm{G}} \end{bmatrix} + 2100$$

$$Z = \left(n_1^{\mathrm{G}}\right)^2 + \left(n_2^{\mathrm{G}}\right)^2 + \left(n_3^{\mathrm{G}}\right)^2 - 20n_1^{\mathrm{G}} - 80n_2^{\mathrm{G}} - 40n_3^{\mathrm{G}} + 10^2 + 40^2 + 20^2$$

$$Z = \left(n_1^{\mathrm{G}}\right)^2 - 2(10)n_1^{\mathrm{G}} + 10^2 + \left(n_2^{\mathrm{G}}\right)^2 - 2(40)n_2^{\mathrm{G}} + 40^2 + \left(n_3^{\mathrm{G}}\right)^2 - 2(20)n_3^{\mathrm{G}} + 20^2$$

$$Z = \left(n_1^{\mathrm{G}} - 10\right)^2 + \left(n_2^{\mathrm{G}} - 40\right)^2 + \left(n_3^{\mathrm{G}} - 20\right)^2$$

**Appendix A-2.** Let $Z = \sum_g \sum_{p \in P_g} \left(n_{g,p}^{\mathrm{GP}} - \hat{n}_{g,p}^{\mathrm{GP}}\right)^2$ . By expansion, $Z = \sum_g \sum_{p \in P_g} \left(n_{g,p}^{\mathrm{GP}}\right)^2 - 2 \sum_g \sum_{p \in P_g} n_{g,p}^{\mathrm{GP}} \hat{n}_{g,p}^{\mathrm{GP}} + \sum_g \sum_{p \in P_g} \left(\hat{n}_{g,p}^{\mathrm{GP}}\right)^2$ .

**Property**: $Z$ is a quadratic function and as such, $Z = \sum_g \left(\frac{1}{2} \left(x_g\right)^{\mathrm{T}} H_g x_g - \Phi_g x_g + c_g\right)$ where:

- $x_g = \begin{bmatrix} n_{g,1}^{\mathrm{GP}} & n_{g,2}^{\mathrm{GP}} & \cdots & n_{g,|P_g|}^{\mathrm{GP}} \end{bmatrix}^{\mathrm{T}}$
- $\Phi_g = \begin{bmatrix} 2\hat{n}_{g,1}^{\mathrm{GP}} & 2\hat{n}_{g,2}^{\mathrm{GP}} & \cdots & 2\hat{n}_{g,|P_g|}^{\mathrm{GP}} \end{bmatrix}$
- $c = \sum_p \left(\hat{n}_{g,p}^{\mathrm{GP}}\right)^2$
- $H_g = 2I_{|P_g|}$ is the Hessian of the function $f = \sum_p \left(n_{g,p}^{\mathrm{GP}}\right)^2, f: \mathbb{R}^{|P_g|} \to \mathbb{R}$.
- Diagonals of $H_g$ are $\frac{\partial^2 f}{\partial n_{g,p}^{\mathrm{GP}}} = 2$ and non-diagonals are $\frac{\partial^2 f}{\partial n_{g,p}^{\mathrm{GP}} \partial n_{g,p'}^{\mathrm{G}}} = 0 \ \forall p, p' | p \neq p'$

**Example:** Given $x_1 = \begin{bmatrix} n_{1,1}^{\mathrm{GP}} & n_{1,2}^{\mathrm{GP}} \end{bmatrix}^{\mathrm{T}}, x_2 = \begin{bmatrix} n_{2,1}^{\mathrm{GP}} \end{bmatrix}$ and $\hat{n}_{1,1}^{\mathrm{GP}} = 10, \hat{n}_{1,2}^{\mathrm{GP}} = 5$ and $\hat{n}_{2,1}^{\mathrm{GP}} = 7$ then:

$$z_1 = \frac{1}{2} (x_1)^{\mathrm{T}} H_1 x_1 - (\Phi_1)^{\mathrm{T}} x_1 + c_1 = \frac{1}{2} \begin{bmatrix} n_{1,1}^{\mathrm{GP}} & n_{1,2}^{\mathrm{GP}} \end{bmatrix} \begin{bmatrix} 2 & 0 \\ 0 & 2 \end{bmatrix} \begin{bmatrix} n_{1,1}^{\mathrm{GP}} \\ n_{1,2}^{\mathrm{GP}} \end{bmatrix} - \begin{bmatrix} 20 & 10 \end{bmatrix} \begin{bmatrix} n_{1,1}^{\mathrm{GP}} \\ n_{1,2}^{\mathrm{GP}} \end{bmatrix} + 125$$

$$z_1 = \left(n_{1,1}^{\mathrm{GP}}\right)^2 - 2(10)n_{1,1}^{\mathrm{GP}} + (10^2) + \left(n_{1,2}^{\mathrm{GP}}\right)^2 - 2(5)n_{1,2}^{\mathrm{GP}} + (5^2) = \left(n_{1,1}^{\mathrm{GP}} - 10\right)^2 + \left(n_{1,2}^{\mathrm{GP}} - 5\right)^2$$

$$z_2 = \frac{1}{2} (x_2)^{\mathrm{T}} H_2 x_2 - (\Phi_2)^{\mathrm{T}} x_2 + c_2 = \frac{1}{2} \left[n_{2,1}^{\mathrm{GP}}\right]^2 - 2(7)n_{2,1}^{\mathrm{GP}} + 7^2 = \left(n_{2,1}^{\mathrm{GP}} - 7\right)^2$$

$$Z = z_1 + z_2 = \left(n_{1,1}^{\mathrm{GP}} - 10\right)^2 + \left(n_{1,2}^{\mathrm{GP}} - 5\right)^2 + \left(n_{2,1}^{\mathrm{GP}} - 7\right)^2$$